\documentclass[useAMS,usenatbib,dvipdfmx]{mn2e}
\usepackage{amsmath}
\usepackage[dvipdfmx]{graphicx}
\usepackage{amsmath,amssymb}
 \usepackage{color}
\newcommand{\msun}{{\rm M}_{\sun}}
\newcommand{\rsun}{{\rm R}_{\sun}}

\title[Pop III binary black holes and the quasi normal mode]
{
The detection rate of Inspiral and Quasi-normal modes of Pop III binary black holes which can confirm or refute the General Relativity in the strong gravity region
}

\author[T. Kinugawa et al.]{Tomoya Kinugawa$^{(1)}$ $ \thanks{E-mail:
kinugawa@tap.scphys.kyoto-u.ac.jp}$, Akinobu Miyamoto$^{(2)}$, Nobuyuki Kanda$^{(2)}$  \and Takashi Nakamura$^{(1)}$\\
\\
$^{1}$Department of Physics, Graduate School of Science, Kyoto University,
Kyoto 606-8502, Japan\\
$^{2}$Department of Physics, Graduate School of Science, Osaka city University,
Osaka, Japan}
\begin{document}

\date{\today}
\maketitle
\begin{abstract}
Using our population synthesis code, we found that the typical chirp mass defined by $(m_1m_2)^{3/5}/(m_1+m_2)^{1/5}$ of Population III (Pop III) binary black holes (BH-BHs)   is $\sim30~\msun$ with the total mass of $\sim60~\msun$  so that the inspiral chirp signal as well as quasi normal mode (QNM) of the merging black hole (BH) are interesting targets of  KAGRA.  The detection rate of the coalescing Pop III BH-BHs  is {$\sim$180 $\rm events~yr^{-1}$$(\rm SFR_p/(10^{-2.5}~\msun \rm~yr^{-1}~Mpc^{-3}))\cdot([f_b/(1+f_b)]/0.33)\cdot Err_{sys}$} in  our standard model where $\rm SFR_{p},~f_b$ and $\rm Err_{sys}$ are the peak value of the Pop III star formation rate, the binary fraction and the  systematic error  with $\rm Err_{sys}=1$ for our standard model, respectively.
To evaluate the robustness of chirp mass distribution and the range of $\rm Err_{sys}$, we examine the dependence of the results on the unknown parameters and the distribution functions in the population synthesis code.
We found that the chirp mass has a peak at  $\sim 30 ~\msun$ in most of  parameters and distribution functions as well as  $\rm Err_{sys}$ ranges from 0.046 to 4.
Therefore, the detection rate of the coalescing Pop III BH-BHs ranges {about $8.3-720\ {\rm events~yr^{-1} ~(SFR_p/(10^{-2.5}~\msun~yr^{-1}~Mpc^{-3}))}\cdot (\rm
[f_b/(1+f_b)]/0.33)$}. 
The minimum rate corresponds to the worst model which we think  unlikely  so that unless $ {\rm ~(SFR_p/(10^{-2.5}~\msun~yr^{-1}~Mpc^{-3}))\cdot([f_b/(1+f_b)]/0.33) \ll 0.1}$, we expect the Pop III BH-BHs merger rate of at least one event per year by  KAGRA.  
{Nakano, Tanaka \& Nakamura (2015) show that if S/N of QNM is larger than 35,  we can confirm or refute the General Relativity (GR) more than 5 sigma level. 
In our standard model, the detection rate of Pop III BH-BHs whose S/N is larger than 35 is $3.2~\rm events~yr^{-1}$$(\rm SFR_p/(10^{-2.5}~\msun \rm~yr^{-1}~Mpc^{-3}))\cdot([f_b/(1+f_b)]/0.33)\cdot Err_{sys}$. 
Thus, there is a good chance to check whether GR is correct or not in the strong gravity region.}

\end{abstract}


\section{Introduction}

The second generation gravitational wave detectors such as KAGRA\footnote{http://gwcenter.icrr.u-tokyo.ac.jp/en/}, Advanced LIGO\footnote{http://www.ligo.caltech.edu/}, Advanced VIRGO\footnote{http://www.ego-gw.it/index.aspx/} and GEO\footnote{http://www.geo600.org/} are under construction and the first detection of gravitational wave is expected in near future.
The most important sources of   gravitational waves  are compact binary mergers such as the binary neutron star (NS-NS), the neutron star black hole binary (NS-BH), and the BH-BH.  As the compact binary radiates gravitational wave and loses the orbital energy and the angular momentum, the compact binary coalesces.
The merger rate of NS-NS can be estimated using the binary pulser observation \citep*[e.g.,][]{Kalogera2004a,Kalogera2004b}.
However  NS-BH and  BH-BH merger rates cannot be estimated using the observation since no such binaries have been observed so that 
they  can be estimated only by the theoretical approach called the population synthesis.
For Population I (Pop I) and Population II (Pop II) stars, the merger rates of compact binaries are  estimated by \cite*{Belczynski2002, Belczynski2007, Belczynski2012a, Dominik2012,Dominik2013}.

%
In this paper, we focus on Pop III stars which were  formed first in the universe with zero metal  after the Big Bang.
The formation process of Pop III stars has been argued by many authors such as \cite*{Omukai1998, Bromm2002, Abel2002, Yoshida2008, Greif2012}.
The simulations of rotating minihalo in the early universe suggest the formation of binaries and multiple star systems \citep*{Machida2008,Stacy2010}.
There are four reasons why we focus on Pop III binaries as gravitational wave sources.

Firstly,  Pop III stars are more massive than Pop I stars \citep*{McKee2008,Hosokawa2011,Hosokawa2012, Stacy2012} so that  Pop III binaries tend to be neutron star (NS) and BH.
Secondly, the merger timescale of compact binaries due to gravitational wave is in proportion to the fourth power of the separation so that even the compact binary 
 formed in the early universe can merge at  present.
Therefore, the Pop III compact binary mergers formed  in the early Universe may be detected by gravitational wave detectors  such as KAGRA.
Thirdly, BH-BH and BH-NS binary formed from Pop III stars tend to be more massive than those formed from Pop I stars while  the detectable distance of the chirp signal is proportional to 5/6 power of the chirp mass defined by $(m_1m_2)^{3/5}/(m_1+m_2)^{1/5}$ of the compact binary \citep{Kinugawa2014}. This means that the detectable distance of Pop III compact binaries is longer than that of Pop I compact binaries so that even if the merger rate  per co-moving volume is smaller, the detection rate can be larger for Pop III BH-BH binaries.  
Fourthly,   Pop III compact binaries as gravitational wave source were considered by \cite*{Belczynski2004}, \cite*{Kowalska2012}, they focused on Pop III stars of mass over hundred solar masses or only on the background.
However the typical mass of Pop III stars is now considered as 10-100 $\rm{M_{\odot}}$ \citep{Hosokawa2011,Hosokawa2012} due to the evaporation of the accretion disk by the strong ultra-violet photons from the central star.
Therefore, Pop III binary population synthesis should be calculated with more realistic lower mass range.

In our previous study \citep{Kinugawa2014}, we  performed  Pop III binary population synthesis with initial mass range of 10-100 $\msun$.
The results showed  that Pop III binaries tend to be massive BH ones with chirp mass   $\sim 30 ~\rm M_{\odot}$.   In our standard model, the detection rate of gravitational wave by the second generation detectors such as KAGRA is $140~ \cdot ({\rm SFR}_{\rm p}/10^{-2.5}\ \msun~{\rm yr}^{-1}~{\rm Mpc}^{-3})\cdot([f_b/(1+f_b)]/0.33) \cdot {\rm Err}_{\rm sys}\ {\rm events}\ {\rm yr}^{-1}$  where $\rm SFR_p,~ f_b$ and $\rm Err_{sys}$ are the peak value of the Pop III star formation rate, the binary fraction\footnote{{The definition of the binary fraction is $N_{binary}/(N_{single}+N_{binary})$ where the $N_{binary}$ and $N_{single}$ are the number of single stars and the number of binary  systems, respectively. Thus, $\rm SFR_p\cdot f_b/(1+f_b)$ means the binary system formation rate. } }
and the possible systematic error due to the assumptions in Pop III Monte Carlo population synthesis, respectively.
${\rm Err_{sys}=1}$ corresponds to  our assumption of binary parameters and initial distribution functions for our standard model. 
However, there are some uncertainties in binary evolution and initial distribution function of Pop III stars so that ${\rm Err_{sys}}$ might not be unity.

Therefore, in this paper, we perform  Pop III binary population synthesis with several binary parameters and initial distribution functions to estimate the variability of merger rates and properties of Pop III compact binaries.  That is, we estimate the possible range of ${\rm Err_{sys}}$ to see if the event rate is larger than ${\rm 1 ~yr^{-1}}$.
 In order to perform Pop III binary evolutions, we renewed Hurley's binary population synthesis code \citep*{Hurley2002} in our previous paper \citep{Kinugawa2014}.
In this paper, we adopt the cosmological parameters of $(\Omega_{\Lambda}, \Omega_{\rm m}) = (0.6825, 0.3175)$, the Hubble parameter of $H_0 = 67.11\ {\rm km}\ {\rm s}^{-1}\ {\rm Mpc}^{-1}$~{and the Hubble time $t_{\rm Hubble}=13.8$ Gyrs }\citep{Planck2013}.

The readers of this journal might not be familiar with the quasi-normal mode of BH in the title of this paper  so that 
we will briefly explain what is the quasi-normal mode of BH here. Let us consider the Schwarzschild BH of mass $M$ and a small particle of mass $m \ll M$ falling into this BH. If the specific angular momentum ($L$) of the particle is smaller than $4GM/c$, the particle spirals into the BH.  At first the gravitational wave  depending on $L$ is emitted but in the final phase the damped gravitational wave is observed as is shown in Figs. 5-4 of \citep{nakamura1987}. The wave form of this damped gravitational wave is expressed by $e^{i(\omega_r+i\omega_i)t}$ where $\omega_r$ and $\omega_i$ depend only on the mass of BH $M$ but not on $L$. Since $\omega_i >0$ in general, this damped oscillation is called the ringing tail or quasi-normal mode. \cite{chandra1975, leaver1985, leaver1986} obtained the complex frequency of quasi-normal modes including Kerr BH case. For the case of Schwarzschild BH, using the WKB approximation, \cite{schutz1985} showed that  $\omega_r$ and $\omega_i$ are determined by the space-time
inside $3GM/c^2$, that is, the space-time near the event horizon  so that the detection of the expected quasi-normal mode of the BH can confirm General Relativity in the strong gravity region. If it is different from the expected value, the true theory of gravity is different from the General Relativity.

This paper is organized as follows.
We describe the models and the calculation method of Pop III binary population synthesis simulations in \S \ref{method}. Results of our calculation such as the properties and the merger rates of Pop III compact binaries  are shown in \S \ref{result}. The discussion \& summary are presented in \S \ref{discussion}.


\section{The method of binary population synthesis simulations}\label{method}
In order to perform Pop III binary evolutions, we use Pop III binary evolution code \citep{Kinugawa2014} which is upgraded from BSE code \footnote{http://astronomy.swin.edu.au/~jhurley/} \citep*{Hurley2002}
for the Pop I stars to Pop III stars case.
The result of binary evolutions depends on binary evolution parameters and initial distribution functions such as common envelope (CE) parameters $\alpha\lambda$, the lose fraction $\beta$ of transferred stellar mass, the natal kick velocity, the initial mass function (IMF), the initial eccentricity distribution function (IEF)   and others  while the details of these parameters  and distribution functions will be shown later.
We perform Monte Carlo simulations using  $10^6$ zero age main sequence (ZAMS) binaries for each model. 
Table \ref{models1} shows the parameters and distribution functions of each  model.
Each column represents the model name, the IMF, the IEF, the natal kick velocity of supernova, the CE parameter $\alpha\lambda$ and the lose fraction $\beta$ of transfer of stellar matter at the Roche lobe over flow (RLOF) in each model. Model name "worst" means the worst combination of the parameter and distribution functions  each model, we calculate 6 cases with different  mass range and the merger criterion at the CE phase so that the total number of the models is 84.

Firstly, we use three mass range cases such as the under100, the over100 and the 140.
In the under100 case, the initial mass range is  $10~\msun\le M \le 100~\msun$.
If the stellar mass becomes over $100~\msun$ by the binary interaction, the binary evolution calculation is stopped because of no fitting formula of the Pop III star evolution for $M>100~\msun$ \citep{Kinugawa2014} due to the lack of numerical data of  the evolution of Pop III star.
This treatment is the same as that in \cite{Kinugawa2014}.
In the over100 case,  the initial mass range is also $10~\msun\le M \le 100~\msun$.
However, in this case, if the stellar mass becomes larger than  $100\msun$ by the binary interaction, the binary evolution calculation continues by extrapolating the  fitting formula of the evolution beyond $100~\msun$.
In the 140 case, the initial mass range is  $10~\msun\le M \le 140~\msun$.
We use the extrapolated fitting formula up to $M=140~\msun$. The reason for up to 140 is that Pop III star of mass larger than 140$\msun$ explodes with no remnant \citep{Heger2003}.

We randomly choose the initial stellar mass from these mass ranges with the initial distribution functions such as the flat IMF, the log flat IMF and the Salpeter IMF (See \S \ref{our standard model} and \ref{IMF model}).  
If the stellar mass is over $140~\msun$ at the supernova, the binary evolution  is stopped because such a high mass star becomes pair instability supernova without compact remnant \citep{Heger2003}.
Numerical simulations of Pop III star formation by
\citep[e.g.,][]{Hosokawa2011,Hosokawa2012, Stacy2012} suggest
that the Pop III protostar grows only up to $\sim$ several $10~\msun$ and the typical mass of Pop III star can be about $40~\msun$.
In recent simulations \citep{Hirano2014,Susa2014}, the typical mass is almost the same as previous study, that is, $40~\msun$, however,  the some Pop III stars can be more massive than $100~\msun$. 
Therefore, we use the initial mass range as $10~\msun\le M \le 140~\msun$ in the 140 case and study the influences of high mass Pop III binaries for the event rate of gravitational wave sources.

Secondly, we use two merger criteria at the CE phase such as the optimistic core-merger criterion and the conservative core-merger criterion \citep{Hurley2002,Belczynski2002,Kinugawa2014}.
In the case of the optimistic core-merger criterion, if the condition $R'_{1}>R'_{\rm L,1}$ or $R'_2>R'_{\rm L,2}$ is fulfilled, the primary star merges with secondary star, where
$R'_{1}$, $R'_{\rm L,1}$, $R'_2$ and $R'_{\rm L,2}$ are the primary stellar radius, the Roche lobe radius of the primary star, the secondary stellar radius and the Roche lobe radius of the secondary star after the CE phase, respectively \citep{Hurley2002}.
On the other hand, the conservative core-merger criterion is $R'_1+R'_2>a_{\rm f}$, where $a_{\rm f}$ is the separation after the CE phase \citep{Belczynski2002}.
\begin{table*}
\caption{The model parameters.
Each column represents the model name, the IMF, the IEF, the natal kick velocity of supernova, the CE parameter $\alpha\lambda$ and the lose fraction $\beta$ of transfer of stellar matter at the RLOF in each model. Model name "worst" means the worst combination of the parameter and distribution functions. }
\label{models1}
\begin{center}
\begin{tabular}{c c c c c c}
\hline
model & IMF & IEF & natal kick velocity ($\rm km~s^{-1}$) & $\alpha\lambda$ & $\beta$\\
\hline
our standard & flat& $e$ & 0 & 1 & function\\
IMF:logflat& $M^{-1}$ & $e$ & 0 & 1 & function\\
IMF:Salpeter& Salpeter & $e$ & 0 & 1 & function\\
IEF:const.& flat& const. & 0 & 1 & function\\
IEF:$e^{-0.5}$& flat& $e^{-0.5}$ & 0 & 1 & function\\
kick 100 $\rm km~s^{-1}$& flat& $e$ & 100 & 1 & function\\
kick 300 $\rm km~s^{-1}$& flat& $e$ & 300 & 1 & function\\
$\alpha\lambda=0.01$& flat& $e$ & 0 & 0.01 & function\\
$\alpha\lambda=0.1$& flat& $e$ & 0 & 0.1 & function\\
$\alpha\lambda=10$& flat& $e$ & 0 & 10 & function\\
$\beta=0$& flat& $e$ & 0 & 1 & 0\\
$\beta=0.5$ & flat& $e$ & 0 & 1 & 0.5\\
$\beta=1$ & flat& $e$ & 0 & 1 & 1\\
Worst & Salpeter & $e^{-0.5}$ & 300 & 0.01 & 1\\
\hline
\end{tabular}
\end{center}
\end{table*}
\subsection{Brief review of Paper I; our standard model}\label{our standard model}

In this paper,  our standard model is the same as the model III.f in \cite{Kinugawa2014}.
In this section, we review the model III.f in \cite{Kinugawa2014} briefly.
The details are shown in \cite{Kinugawa2014}. 
In order to simulate the binary evolution, we need to choose initial binary parameters such as the primary mass, the mass ratio, the separation and the eccentricity.
These parameters are decided randomly by the initial distribution function and the Monte Carlo method.
In our standard  model, we adopt the flat initial distribution function for the primary mass, the flat function for the mass ratio $f(q)\propto \rm const.$ \citep{Kobulnicky2007,Kobulnicky2012}, the log flat function for the separation $f(a)\propto 1/a$ \citep{Abt1983} and the thermal equilibrium distribution function of the eccentricity $\propto e$ \citep{Heggie1975,Duquennoy1991}.
We use these initial distribution functions except the IMF referenced by the Pop I stars observations because there are no observations suggestions of Pop III binaries initial distribution functions.
As for the IMF, some simulations suggest the flat or the log flat IMF \citep*{Hirano2014, Susa2014}.  
Using these initial distribution functions, we put the ZAMS binary and start the evolution of the binary.
In order to calculate each stellar evolution, we use the formula fitted to the numerical calculations of Pop III stellar evolutions by  \cite{Marigo2001}.
This fitting formula is described by  the stellar radius and the core mass as a function of the stellar mass and the time from the birth of each star.
The details of fitting equations for Pop III are shown in \cite{Kinugawa2014}.
We can calculate the  evolution the binary adding the binary interactions to the Pop III star evolution using the fitting formula.
We also need to consider the binary interactions such as the tidal friction, the Roche lobe over flow, the CE phase, the effect of the supernova explosion and the back reaction  of the gravitational wave. 

Firstly, we review the tidal friction.
The tidal force from the companion star changes the stellar radius and the shape.
In general, the stellar spin angular velocity is  different from the orbital angular velocity.
Therefore  the vector of the tidal deformation is different from the vector of the tidal force so that  the tidal torque is generated.
The tidal torque transfers the angular momentum from the stellar spin to the orbital angular momentum. 
This interaction changes the binary separation, eccentricity and the  spins of each star.

Secondly, we review the RLOF.
When the star evolves and the stellar radius becomes large, the stellar matter is captured by  the companion star and is transferred to the companion star.
This phenomenon is called as the RLOF.
We call the donor star as the primary star.
The recipient star is called as the secondary star. 
The region within which the stellar material is gravitationally bound to the star is called as the Roche lobe radius of that star.
If the primary stellar radius becomes larger than the Roche lobe radius, the primary stellar matter migrates to the secondary star.
We define a coefficient $\beta$ as the lose fraction of transferred stellar mass descrived as  
\begin{equation}\label{betaMT}
\dot{M_2}=-(1-\beta)\dot{M_1}.
\end{equation}
In this case, the accretion rate to the secondary star and $\beta$ is determined by the method of \cite{Hurley2002}.
If the secondary star is in the main sequence phase  or in the He-burning phase, we assume the accretion rate is expressed by
\begin{equation}\label{MT}
\dot{M}_{2}=-{\rm{min}}\left(10\frac{\tau_{\dot{M}}}{\tau_{{\rm{KH,2}}}},1\right)\dot{M}_{1},
\end{equation}
where $\dot{M_1}$ is the mass loss rate of the primary star and $\tau_{\dot{M}}$ is the accretion time scale defined as
\begin{equation}
\tau_{\dot{M}}\equiv\frac{M_2}{\dot{M}_1},
\end{equation}
 The Kelvin-Helmholtz timescale $\tau_{\rm KH,2}$ is defined as
\begin{align}
\tau_{\rm{KH,2}}
=\frac{GM_2(M_2-M_{\rm{c},2})}{L_2R_2},
\end{align}
where $M_2,~M_{\rm{c},2},~L_2$ and $R_2$ are the secondary stellar mass, the core mass, the luminosity and the radius of the star, respectively.
If the secondary star is in the He-shell burning phase, we assume the secondary star can get  all matter of the primary.
Thus,
\begin{equation}\label{MT2}
\dot{M}_{2}=-\dot{M}_{1}.
\end{equation}
In our standard model, we use $\beta$ function defined by Eq.(2) which is computed by the  fitting formulae by \cite{Hurley2002}. However, the accretion rate of the secondary which is not a compact object,  is not  understood well 
so that we use also the accretion rate of the secondary  described by the constant $\beta$ parameter (For details see \S \ref{RLOF}).  
  
If the secondary star is a compact object such as NS and BH, we always use $\beta=0$ and
the maximum of the accretion rate is limited by the Eddington mass accretion rate defined by
 \begin{align}
\dot{M}_{\rm{Edd}}&=\frac{4\pi c R_2}{\kappa_{\rm T}}\\\nonumber
&=2.08\times10^{-3}(1+X)^{-1}
\left(\frac{R_2}{\rsun}\right)~\rm{M_{\odot}~yr^{-1}},
\end{align}
where $\kappa_{\rm T} =0.2(1+X)\ \rm cm^2\ g^{-1}$ is the Thomson scattering opacity and $X(=0.76)$ is the H-mass fraction for Pop III star. 

When the primary star is a giant and the mass transfer is dynamically unstable, the secondary star plunges into the primary envelope and the binary enters the CE phase.
In this phase, the friction between the primary star and the secondary star yields the loss of  the angular momentum of the secondary to decrease  the binary separation,
 while the  envelope of the primary is evaporated by the energy liberated through the friction.
Consequently, the binary either becomes the close binary or merges during the CE phase.
Now we define $a_{\rm{i}},~a_{\rm{f}},~M_1,~M_{\rm c,1},~M_{\rm{env,1}}$ and $R_1$ as the separation before the CE phase, the separation after the CE phase, the primary mass, the primary core mass, the primary envelope mass and the primary separation, respectively.
The separation after the CE phase is calculated by the energy formalism \citep{Webbink1984} defined by
\begin{equation}
\alpha\left(\frac{GM_{\rm{c,1}}M_2}{2a_{\rm{f}}}-\frac{GM_1M_2}{2a_{\rm{i}}}\right)=\frac{GM_{\rm{1}}M_{\rm{env,1}}}{\lambda R_1}, \label{eq:ce2}
\end{equation} 
where $\alpha$ and $\lambda$ are the efficiency and the binding energy parameter, respectively.
In our standard model, we adopt $\alpha\lambda=1$.

In our calculation, we adopt two merger criteria  during the CE phase
which is shown already in the last paragraph of \S 2 before \S 2.1.

When the supernova explosion  occurs, the sudden mass ejection and the natal kick make the binary obit to change drastically.
In our standard model, we adopt the natal kick velocity equal to zero.
In this case, the binary orbit changes  only by the mass ejection effect in the supernova event.
The separation and the eccentricity after the supernova explosion  are described as
\begin{equation}
a'=\left(\frac{|\bf{v}|^2}{GM_{\rm{total}}}-\frac{|\bf v|^2}{GM_{\rm{total}}'}+\frac{1}{a}\right)^{-1}\label{eq:an},
\end{equation}
\begin{equation}
e'=\sqrt{1-\frac{|\mathbf{r}\times\mathbf{v}|^2}{GM'_{\rm{total}}a'}},\label{eq:en}.
\end{equation}
where $M_{\rm total}$ is the total mass,
the superscript $'$ means the value after the supernova while 
$v$, $\bf v$ and $\bf r$ are the relative speed, the relative velocity and the separation vector before the supernova, respectively \citep*{Blaauw1961}.
  
After the binary becomes the compact binary due to above binary interactions, the orbit of the binary  shrinks
by the emission of the gravitational waves.
The separation and the eccentricity are given by
\begin{equation}
\frac{\dot{a}}{a}=-\frac{64G^3M_1M_2M_{\rm{total}}}{5c^5a^4}\frac{1+\frac{73}{24}e^2+\frac{37}{96}e^4}{(1-e^2)^{7/2}},\label{eq:semimajor}
\end{equation}
\begin{equation}
\frac{\dot{e}}{e}=-\frac{304G^3M_1M_2M_{\rm{total}}}{15c^5a^4}\frac{1+\frac{121}{304}e^2}{(1-e^2)^{5/2}}\label{eq:edot}
\end{equation} 
\citep*{Peters1963,Peters1964}.
We calculate the binary evolutions taking account all  these binary interactions and estimate how many binaries become the compact binaries which merge within the Hubble time.

\subsection{Parameter Surveys}
To estimate the range of $\rm Err_{sys}$, we calculate the binary population synthesis using other initial distribution functions and  binary parameters since for Pop III stars
we have no information on these functions and  parameters. 
However  in this paper we do not take into account the dependence on the initial separation function and the initial mass ratio function
because unlike an initial eccentricity functions, there are no suggestions for other distribution functions for massive binaries although 
possible  dependence of $\rm Err_{sys}$ on the change of these initial distribution functions is discussed in \S 4 (Discussion).
   
\subsubsection{log flat IMF  and Salpeter models}\label{IMF model}
These models correspond to  the log flat IMF, that is IMF$\propto d\log (M)$, and the Salpeter IMF.
In recent numerical simulations \citep{Hirano2014, Susa2014}, IMF of Pop III stars might be  the log flat IMF.
On the other hand, Salpeter IMF is acceptable as Pop I IMF.
We calculate these IMF models in order to estimate the IMF dependence.  
The other initial distribution functions and binary parameters are the same as in our standard model.

\subsubsection{IEF:const. and $ e^{-0.5}$ models}
In these models, the IEF are changed from our standard model.
In general  the initial eccentricity distribution might be  the thermal-equilibrium distribution (IEF:$2e$) \citep{Heggie1975}.
However in recent observation of massive binaries, the eccentricity distribution is not the thermal-equilibrium distribution. 
The observation of massive multiple-star systems in the Cygnus OB2 \citep{Kobulnicky2014} implies that the observed  IEF is consistent with uniform one.
On the other hand, the observation of massive binaries ($M>15~\msun$) \citep{Sana2012} suggests that the power law for the distribution function of eccentricity as $propto e^{-0.5}$. 
Thus, we calculate these two initial eccentricity distribution function models.
The other initial distribution functions and binary parameters are the same as in our standard model.

\subsubsection{kick 100 $ km~s^{-1}$ and kick 300 $km~s^{-1}$ models}
The pulsar observations suggest the existence of the NS kick.
It is observed that the young NSs move with velocities in range of $200-500~\rm km~s^{-1}$  \citep[e.g.][]{Lyne1994, Hansen1997}.
Since the NS kick velocity either disrupts the binary or increases the separation, the formation rate and the coalescing time of the NS-NS and the NS-BH  depend on the NS kick velocity.  
On the other hand, the formation rate and the coalescing time of the BH-BH might have nothing to do with the natal kick velocity because the BH progenitor directly collapses to the BH.  However, \cite{Repetto2012} suggests  that the stellar mass BHs have the natal kicks comparable to NSs from  the distance distribution of the Galactic BH (low mass X-ray binaries)  above the  galactic plane. Pop III BH-BHs are massive, so that they may not have  such natal kick as stellar mass BHs.
However there  is no observation of Pop III BHs. Thus we cannot definitely claim that Pop III BHs do not have natal kicks.
Therefore, we  take into account  the natal kick for both NS and BH.
In order to estimate the dependence on the natal kick, we calculate two models.
In these models, when stars become compact objects such as NS and BH, we assume that  the natal kick speed $v_{\rm k}$ obeys a isotropic Maxwellian distribution as
\begin{equation}
P(v_{\rm k})=\sqrt{\frac{2}{\pi}}\frac{v_{\rm k}^2}{\sigma_{\rm k}^2}\exp\left[-\frac{v_{\rm k}^2}{\sigma_{\rm k}^2}\right],
\end{equation}
where $\sigma_{\rm k}$ is the dispersion.
In the kick 100 $\rm km~s^{-1}$ model and kick 300 $\rm km~s^{-1}$ models, we uses $\sigma_{\rm k}=100 ~\rm km~s^{-1}$ and $\sigma_{\rm k}=300 ~\rm km~s^{-1}$, respectively.
The details of the method how to calculate the natal kick are shown  in \cite{Hurley2002}.
The other initial distribution functions and binary parameters are the same as in our standard model. 

\subsubsection{$\alpha\lambda=0.01$, $\alpha\lambda=0.1$ and $\alpha\lambda=10$ models}
If the primary star becomes a giant and it begins dynamically unstable mass transfer so that the secondary star can be engulfed into the envelope of the primary star. In such case, the binary enters the CE phase. 
Once the secondary star is swallowed up by the envelope of the primary star, it spirals into the core of the primary star due to the orbital energy and angular momentum loss by the friction.
It is assumed that this spiral-in continues until  all the envelope of the primary star  is ejected from the binary system. 
The separation after the CE phase $a_{\rm f}$ is calculated using CE energy balance prescription of Eq.  \ref{eq:ce2}

However CE parameters are uncertain.
In general, it is  assumed that $\alpha\lambda=1$ since only $\alpha\lambda$ is the meaningful parameter.
Our  standard model uses  $\alpha\lambda=1$.
However, these parameters can be other values.
Therefore, we calculate three  cases of the CE parameters as $\alpha\lambda=0.01,~0.1,~10$.
The other initial distribution functions and binary parameters are the same as in our standard model.

\subsubsection{$\beta=0$, $\beta=0.5$ and $\beta=1$ models}\label{RLOF}
If the primary star fulfills Roche lobe and it begins dynamically stable mass transfer, the secondary gets the mass from the primary star.
$\beta$ is called as the lose fraction of transferred stellar mass defined as Eq. \ref{betaMT}.
It is considered that $\beta$ varies depending on the binary \citep*[e.g.][]{Eggleton2000}.
In binary population synthesis study, $\beta$ is treated as a function or the constant parameter.
In our standard model, we use $\beta$ as a function (See section \ref{our standard model}).
On the other hand, in other studies $\beta$ is treated as the constant \citep[e.g.][]{Belczynski2002}.
Thus, we estimate the variabilities of result for  three constant $\beta$ cases.
The other initial distribution functions and binary parameters are the same as in our standard model. 
Furthermore, if the mass transfer is nonconservative ($\beta>0$), the criterion of the stability of the mass transfer should be changed from the criterion of \citep{Kinugawa2014,Hurley2002} because this criterion assumes that the mass transfer is conservative.
We use the criterion of \cite{Eggleton2011} as
\begin{align}
\zeta_{\rm L}&=\frac{d{\rm{log}}R_{\rm L,1}}{d{\rm{log}}M_1}\notag\\
                  &=\frac{(0.33+0.13q_1)(1+q_1-\beta q_1)+(1-\beta)(q_1^2-1)-\beta q_1}{1+q_1},
\end{align}   
where $R_{\rm L,1}$ and $q_1=M_1/M_2$ are the Roche lobe radius and the mass ratio.
If $\zeta_{\rm ad}={d{\rm{log}}R_{{\rm ad},1}}/{d{\rm{log}}M_1}<\zeta_{\rm L}$ where $R_{\rm ad}$ is the radius when the star reaches hydrostatic equilibrium, the binary starts the dynamically unstable mass transfer.  

\subsubsection{Worst model}
In this model, we adopt  the initial conditions and binary parameters which make the worst result in IMF, IEF, kick, $\alpha\lambda$ and $\beta$ (See Section \ref{result}).
Namely we adopt IMF:Salpeter, IEF:$e^{-0.5}$, kick 300 $\rm km~s^{-1}$, $\alpha\lambda=0.01$ and $\beta=1$.
We, however, think that this worst case is unlikely so that the worst case will teach us the minimum merging rate of Pop III BH-BHs.
{Note  that other combination of parameters may yield even lower rates.
However, we cannot calculate and check all $3\times 3\times3\times4\times4=432$ models.
Thus, we choose the worst parameters from each parameter region.}

\section{Results}\label{result}
\subsection{The properties of Pop III compact binaries}\label{properties}
In order to study the property of Pop III compact binaries, we now show the number of the compact binary formations, the number of the compact binaries which merge within 15 Gyrs and the distribution of the BH-BH chirp mass.
The details of the difference between each model will be shown  in  the following sub-sections (a) to (g).  

Tables \ref{model1} to  \ref{model14} show  the numbers of NS-NS, NS-BH and BH-BH binaries for each model from the initial $10^6$ zero age main star binary. The meanings of  under 100, over 100 and 140 are explained 
 in the first paragraph of \S 2.  The title of each table  comes from the change of some parameter or that of IMF or IEF  from our standard model. In each table we also tabulated  the number of the compact binaries which merge within 15 Gyrs.
The numbers in the parenthesis  are for the case of the 
conservative core-merger criterion while those without the parenthesis are for the case of the optimistic 
core-merger criterion. The meanings of these two criteria were explained in the last paragraph of \S 2 before \S 2.1

In most cases, the number of merging NS-NSs and merging NS-BH are very small or zero as one can notice easily from  Tables 2 to 14. The reasons are as follows.
The wind mass loss and the mass loss by the binary interactions are not so effective for the Pop III binaries because of  the zero metallicity  and smaller radius of Pop III stars   so that the Pop III binaries tend to disrupt or to increase the separation  by the supernova mass ejection \citep{Kinugawa2014}.
Therefore  we focus on the  description of  the  BH-BHs.
Figs. \ref{our standardmass} to  \ref{worstmass} show the chirp mass distribution of BH-BH binaries which merge within 15Gyr for each model. 
In each figure, the red, green, blue, pink, light blue and grey  lines correspond to under100 case with optimistic core-merger criterion,  over100 case with optimistic core-merger criterion,  140 case with optimistic core-merger criterion, under100 case with conservative core-merger criterion,  over100 case with conservative core-merger criterion and  140 case with conservative core-merger criterion, respectively. One can see that in all models, the peak of the observable chirp mass distribution  is about $30~\msun$. Pop III binaries with each mass $M<50~\msun$ are unlikely to be the CE phase.
They evolve via some mass transfer phases and their mass loss is smaller than the evolution passes via a CE phase.
They tend to lose 1/10-1/3 of their mass so that they tend to be 20-30 $\msun$ BH-BHs.   Pop III binaries with each mass $M > 50~\msun$ are likely to be the CE phase and they lose  1/2-2/3 of their mass so that they tend to be 25-30 $\msun$ BH-BHs too.  Therefore, the peak of chirp mass become 25-30 $\msun$.

Figs. \ref{IMFtime} to  \ref{saiakutime} show the dependence of merger time distribution of BH-BH binaries for each model.
In each figure, we describe under100 cases with optimistic core-merger criterion because the merger time distributions do not change a lot in other cases and core-merger criteria.  The most important characteristic is that the merger rate for $t > 10^{9.5}$yr is almost constant for every model.

\subsubsection*{(a)our standard model}
\begin{figure}

  \includegraphics[scale=0.6]{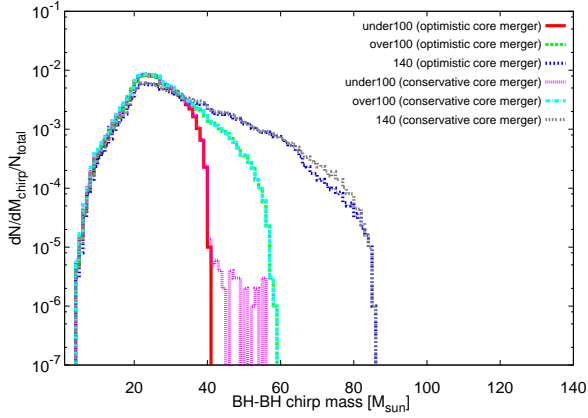} 
\smallskip

\caption{our standard model}

Each line is the normalized distribution of the BH-BH chirp mass.
The red, green, blue, pink, light blue and grey  lines are the under100 case with optimistic core-merger criterion, the over100 case with optimistic core-merger criterion, the 140 case with optimistic core-merger criterion, the under100 case with conservative core-merger criterion, the over100 case with conservative core-merger criterion and the 140 case with conservative core-merger criterion, respectively. $\rm N_{total}=10^6$ binaries.

\label{our standardmass}
\end{figure}

Under100 case is the same as the result of the model III.f in \cite{Kinugawa2014}.
Over100 case is equal to the result of under100 case plus the binaries whose star becomes more massive than $100~\msun$ by the mass transfer so that the number of  massive BH-BH mergers in over100 case is more than that of the under100 case (See Fig. \ref{our standardmass}).
Thus, the number of Pop III BH-BHs which merge within 15 Gyrs in  over100 case increases about ten percent compared to the that of under100 case (See Table \ref{model1}).  
However, the peak of the chirp mass is not changed (See Fig. \ref{our standardmass})
In  140 case the peak of the chirp mass distribution (Fig. \ref{our standardmass}) is almost the same as that of under100 and over100 cases, but the chirp mass distribution of 140 case has tail in the region of  25-60 $\msun$ since the high mass Pop III binaries ($>100~\msun$) tend to be high mass BH-BHs (40-60 $\msun$) via the CE phase.  The event rate of high chirp mass BH-BHs is large  because  the detectable volume (V) is $V\propto M_{\rm chirp}^{5/2}$.
The tail of chirp mass distribution $dN/dM$ in 140 case is proportional to $M^{-5/2}$ in the range of 25-60 $\msun$.

\subsubsection*{(b) log flat  and Salpeter IMF models}
\begin{figure}
  
  \includegraphics[scale=0.6]{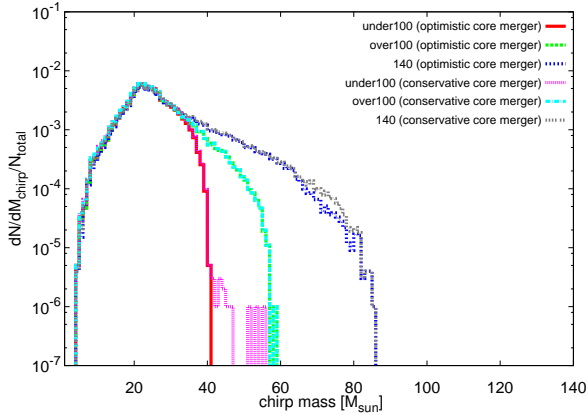} 
\smallskip

\caption{IMF:logflat
}

Same as Fig.\ref{our standardmass} but for IMF:logflat model.

\label{imfmass}
\end{figure}

\begin{figure}
 
  \includegraphics[scale=0.6]{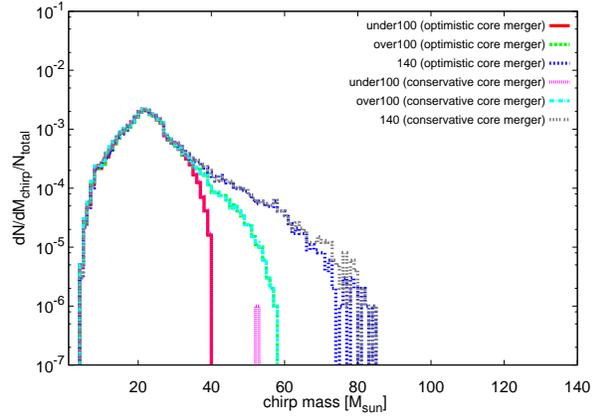} 
\smallskip

\caption{IMF:Salpeter
}

Same as Fig.\ref{our standardmass} but for IMF:Salpeter model.

\label{salmass}
\end{figure}

In these models, the initial masses tend to be smaller than that of our standard model.
Thus, the numbers of the BH-BH formations and merging BH-BHs decrease because the BH-BH progenitors decrease due to IMFs (See Table \ref{model2} and \ref{model3}).
{On the properties of the chirp mass distributions in the log flat and Salpeter IMF models, the number of merging BH-BHs where each mass is more massive than $30~\msun$ is smaller  than our standard model due to the steepness of IMFs.
However, the peak mass is independent on the IMF (See Figs. \ref{imfmass} and \ref{salmass}).}
In 140 cases, the chirp mass distribution $dN/dM$ of IMF for log flat model is proportional to $M^{-3}$ and that of Salpeter model is proportional to $M^{-3.9}$ in a range of 25-60 $\msun$. 
{In  Fig. \ref{IMFtime}, the shape of each merger time distribution is almost the same independent of IMFs, although the number of BH-BHs decreases dependent on the steepness of IMF}.

\subsubsection*{(c)IEF:const. and $e^{-0.5}$ models}
\begin{figure}
 
  \includegraphics[scale=0.6]{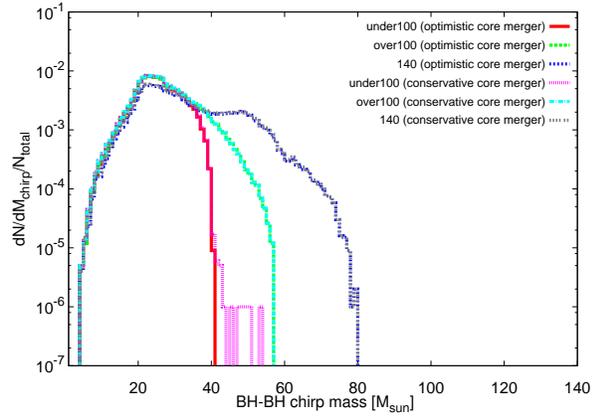} 
\smallskip

  \hspace{2.5pc}

\caption{IEF:const.
}

Same as Fig.\ref{our standardmass} but for IEF:const. model.

\label{eccmass1}
\end{figure}

\begin{figure}
  
  \includegraphics[scale=0.6]{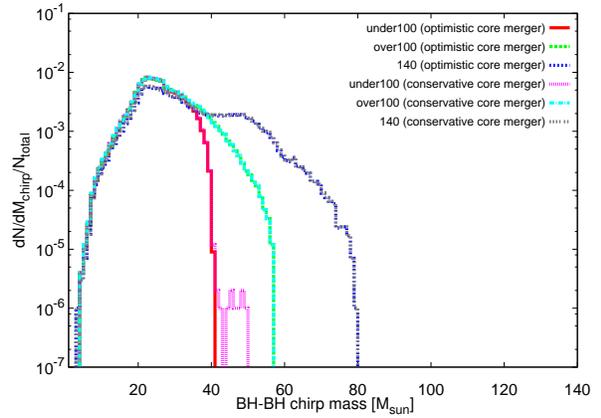} 
\smallskip

\caption{IEF:$e^{-0.5}.$
}

Same as Fig.\ref{our standardmass} but for IEF:$e^{-0.5}$ model.

\label{eccmass2}
\end{figure}

In these models, the initial eccentricities tend to be smaller than that of our standard model. 
If the initial eccentricity is small, the decrease of the separation by binary interactions such as the tidal friction and the gravitational radiation is suppressed.
Thus, the merger rate by the binary interaction decreases.
Therefore, the number of BH-BH formation rate  increases while  the number of merging BH-BHs deceases although the influence to the merger rate is very small (See Table \ref{model4} and \ref{model5}).
The properties of the chirp mass distributions are almost the same as in our standard model (See Fig. \ref{eccmass1}, \ref{eccmass2}).
The properties of the merger time distributions are almost the same as in our standard model too (See Fig. \ref{ecctime}).
We can say that the effect of the eccentricity distribution is  not so large.

\subsubsection*{(d)kick 100 $\rm km~s^{-1}$ and kick 300 $\rm km~s^{-1}$ models}
\begin{figure}
  \includegraphics[scale=0.6]{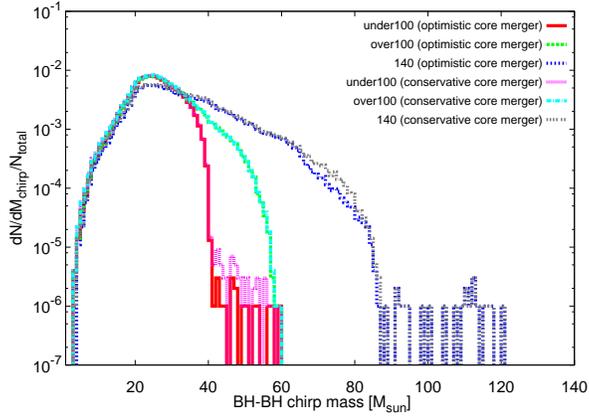} 
\smallskip

\caption{kick 100 $\rm km~s^{-1}$
}

Same as Fig.\ref{our standardmass} but for kick 100 $\rm km~s^{-1}$ model.

\label{kickmass1}
\end{figure}

{In these models, the natal kick disrupts binaries or makes their orbits wide and eccentric.}
Firstly, we argue the kick 100 $\rm km~s^{-1}$ model.
The number of the BH-BH formations decreases by the natal kick compared to our standard model.
However, the number of merging BH-BHs becomes about 1.5 times larger than  that of our standard model (See Table \ref{model6}). 
{The merger timescale by the gravitational radiation is given by
.\begin{align}
T_{\rm merge}(e_0=0)=\frac{5}{256}\frac{a_0^4}{c}\left(\frac{GM_1}{c^2}\right)^{-1}\left(\frac{GM_2}{c^2}\right)\left(\frac{GM_{\rm total}}{c^2}\right)\notag\\
=10^{10}{\rm yr}\left(\frac{a_0}{43~{\rm R_{\odot}}}\right)^{4}\left(\frac{M_1}{30~\msun}\right)^{-1}\left(\frac{M_2}{30~\msun}\right)^{-1}\left(\frac{M_{\rm total}}{60~\msun}\right)^{-1} 
\end{align}
where $a_0$ and $e_0$ are the separation and the eccentricity when the BH-BH is formed.
Thus, the BH-BHs whose separation is larger than about $50~\rm R_{\odot}$ cannot merge within Hubble time. 
On the other hand, the escape velocity and the orbital velocity of the binary system are given by
\begin{align}
v_{\rm esc}&=\sqrt{\frac{2GM}{a}}\\
               &=500~{\rm km~s^{-1}}\left(\frac{M}{30~\msun}\right)^{1/2}\left(\frac{a}{43~\rm R_{\odot}}\right)^{-1/2},
\end{align}
\begin{align}
v_{\rm orb}&=\sqrt{\frac{GM}{a}}\\
               &=350~{\rm km~s^{-1}}\left(\frac{M}{30~\msun}\right)^{1/2}\left(\frac{a}{43~\rm R_{\odot}}\right)^{-1/2},
\end{align}
where $a$ is the separation.
 Equations 15-18 tells us that even though the kick velocity which is 100 $\rm km~s^{-1}$ is aligned with the orbital velocity, the BH-BH progenitors whose separation is smaller than $43~\rm R_{\odot}$ cannot be disrupted by the natal kick.}
Therefore, the BH-BHs disrupted by the natal kick of  100 $\rm km~s^{-1}$ rarely contribute to the merger rate of Pop III BH-BHs from the beginning.
While the increase of the eccentricity by the natal kick reduces the merger timescale due to the gravitational  waves since $T_{\rm merge}(e)\sim (1-e^2)^{7/2}T_{\rm merge}(e_0=0)$ \citep{Peters1963, Peters1964},
 so the number of merging BH-BHs  is larger than that of our standard model. 
The chirp mass distribution is, however, almost the same as that of our standard model (See Fig. \ref{kickmass1}).
In Fig. \ref{kicktime}, the merger time distribution of the kick 100 $\rm km~s^{-1}$ model is almost the same as in our standard model.
However, the number of merging BH-BHs which merges at the early universe is larger than that of our standard model due to the increase of the eccentricity by the natal kick.

\begin{figure}

  \includegraphics[scale=0.6]{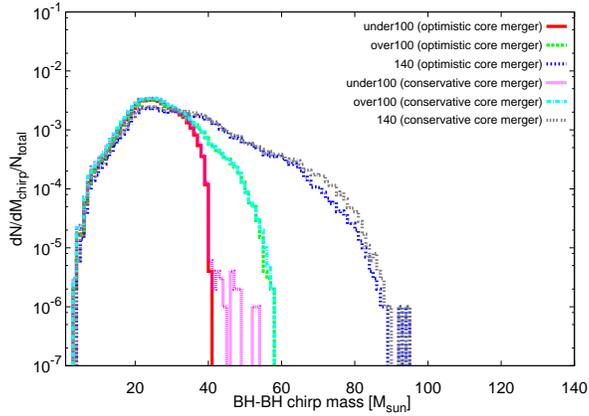} 
\smallskip

\caption{kick 300 $\rm km~s^{-1}$
}

Same as Fig.\ref{our standardmass} but for kick 300 $\rm km~s^{-1}$ model.

\label{kickmass2}
\end{figure}

Secondly, we argue the kick 300 $\rm km~s^{-1}$ model.
The number of the BH-BH formations decreases by the natal kick compared to our standard model and the kick 100 $\rm km~s^{-1}$ model.
The number of the merging BH-BHs also decreases unlike the kick 100 $\rm km~s^{-1}$ model.
{The reasons are that the sum of the orbital velocity and the kick velocity sometimes can exceed $500~ \rm km~s^{-1}$ and that if the binary was not disrupted, the separation of compact binaries tends to extremely widened by the high natal kick velocity.
Thus, the many binaries will be disrupted and the coalescence time of the survived binaries due to the emission of the gravitational wave tends to be so long that the Pop III binaries cannot merge within 15 Gyrs.}
Especially, the low mass binaries are more susceptible to this effect than the massive binaries.
The chirp mass distributions is also  changed and the peak mass region moves to a little more massive than that of our standard model (See Fig. \ref{kickmass2}).
In Fig. \ref{kicktime}, the merger time distribution of the kick 300 $\rm km~s^{-1}$ model is different compared to our standard model.
The number of BH-BHs which merge at the early universe is larger than that of our standard model due to the increase of the eccentricity by the natal kick.  
On the other hand, the progenitors of BH-BHs which have large separation but can merge within the Hubble time in our standard model tend to be disrupted by the natal kick.
Thus, the number of merging  BH-BHs which have large merger time decrease. 

\subsubsection*{(e)$\alpha\lambda=0.01$, $\alpha\lambda=0.1$ and $\alpha\lambda=10$ models}
In these models,  the CE phase results in the merger of binary or the  change of the separation.
The separation after the CE phase is determined by the CE parameter $\alpha\lambda$.
If $\alpha\lambda$ is small, the separation after the CE phase tends to be small and to merge during the CE phase due to the increase in the loss of the orbital energy.
Easily  the binary does not survive during the CE phase.
But if binary survives during the CE phase, the binary  easily merges by the large luminosity of  GW  emission due to the tight orbit.  
On the other hand, if $\alpha\lambda$ is  large, the separation after the CE phase tends to be large due to the decrease in the loss of the orbital energy.
Thus, the binary tends to survive during the CE phase.
When the binary survives in the CE phase, however,  it is  hard to merge due to the small  luminosity of GW.  

\begin{figure}
  \includegraphics[scale=0.6]{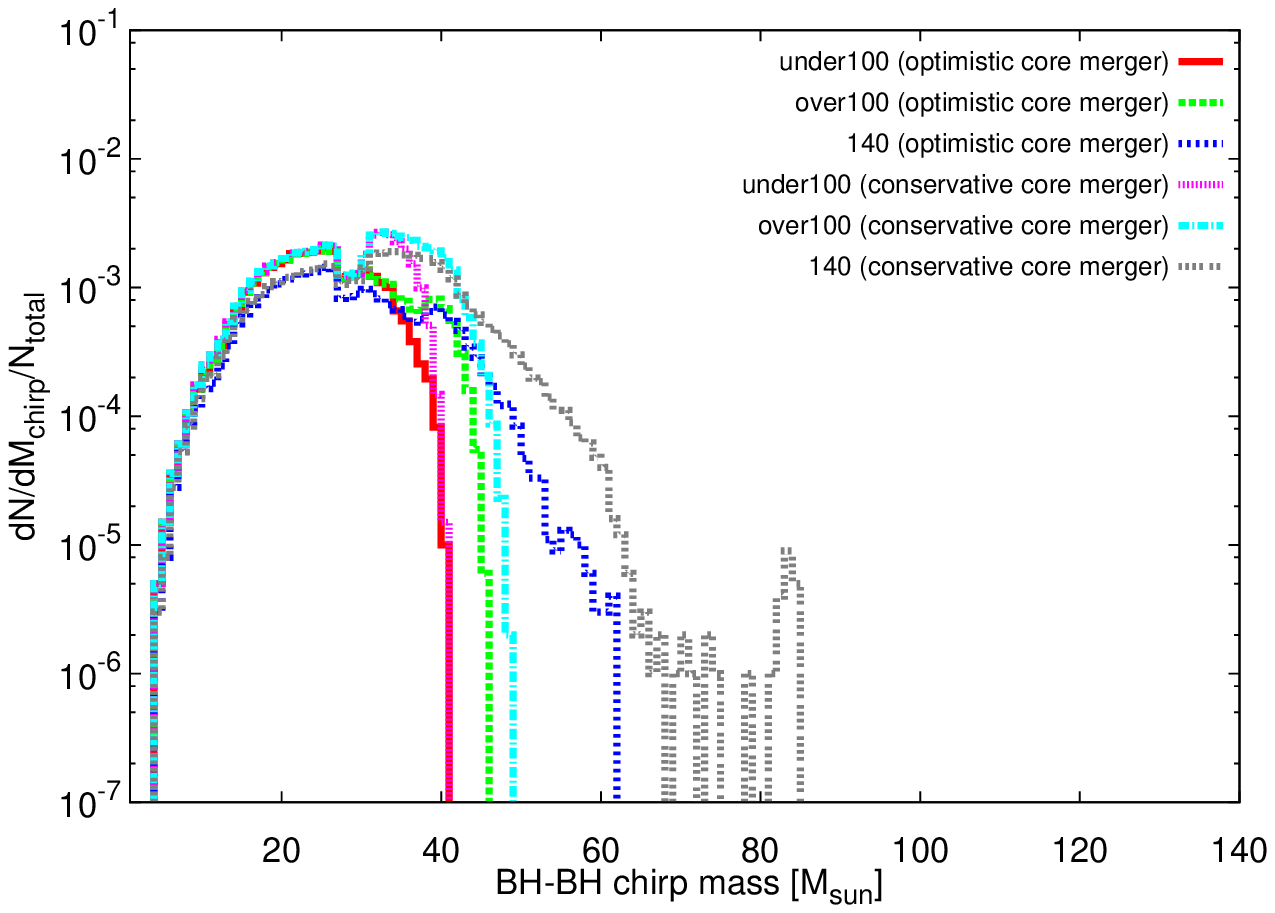} 
\smallskip

\caption{$\alpha\lambda =0.01$
}

Same as Fig.\ref{our standardmass} but for $\alpha\lambda=0.01$ model.

\label{almass1}
\end{figure}

Firstly, we consider  $\alpha\lambda=0.01$ model.
In this case, the number of the BH-BH formations decreases and the number of the merging BH-BHs becomes about 1/3 as large as  our standard model.
{The parameter $\alpha\lambda$ is so small that the almost all binaries merge during the CE phase.}
The Pop III giant with  initial mass  larger than 50 $\rm M_{\odot}$ can enter the CE phase and these binaries tend to  merge.
Thus, the survived binaries evolved via the RLOF.
The binaries which evolved via the RLOF tend to be smaller than 50 $\msun$ and the massive binaries merge more easily.
Thus, in the chirp mass distribution (Fig. \ref{almass1}), the number of the merging high mass BH-BHs becomes smaller than our standard model.
Especially,  140 case is easily affected by these effects. 
In Fig. \ref{altime}, the shortest merger time is larger than that of our standard model, {because almost all progenitors of BH-BHs merge during the CE phase.}

\begin{figure}
  \includegraphics[scale=0.6]{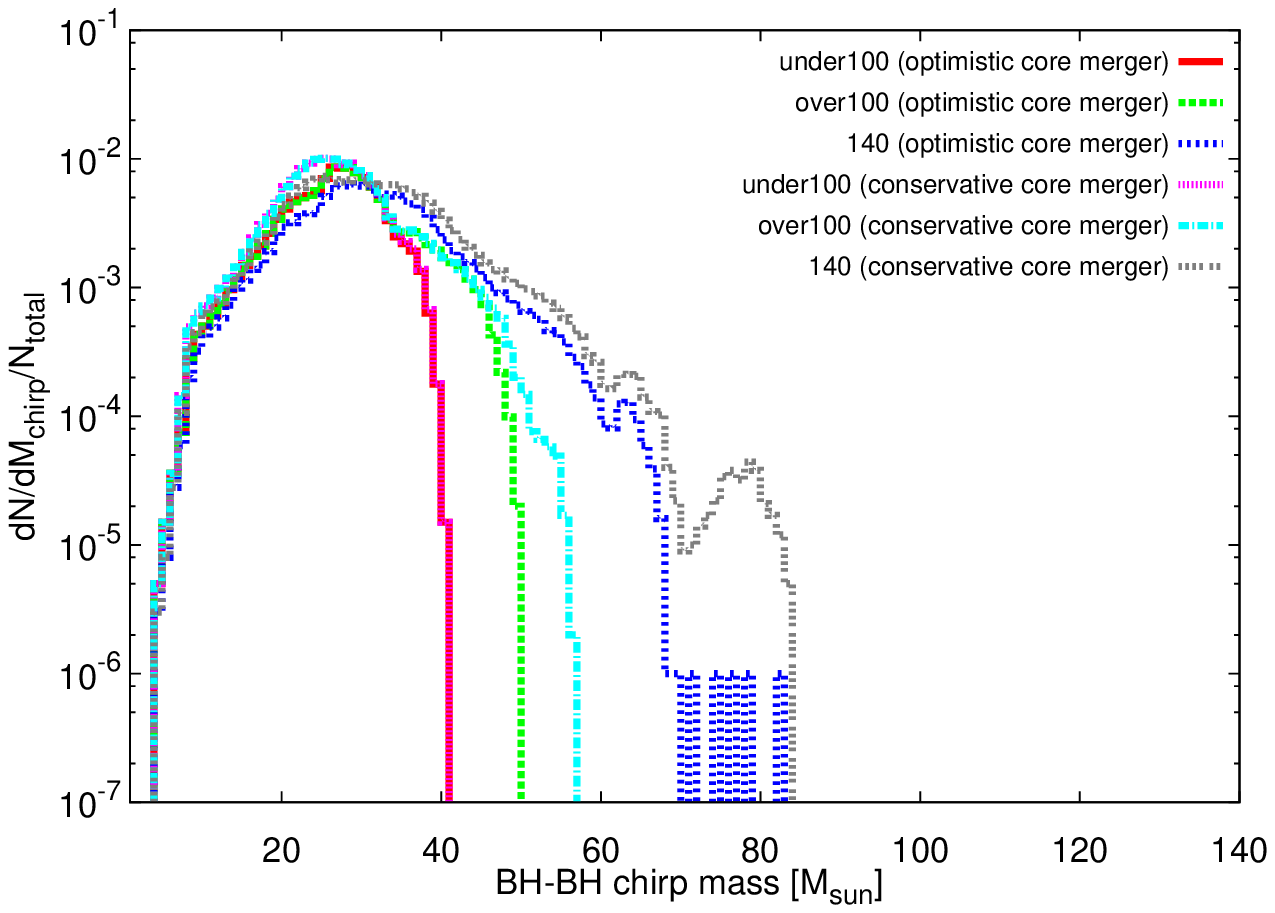} 
\smallskip

\caption{$\alpha\lambda =0.1$
}

Same as Fig.\ref{our standardmass} but for $\alpha\lambda=0.1$ model.

\label{almass2}
\end{figure}

Secondly, we consider the $\alpha\lambda =0.1$ model.
In this case, the number of the BH-BH formation decreases compared with our standard model.
While, the number of the merging BH-BHs is almost the same as that of  our standard model.
Owing to the small $\alpha\lambda$, the binary does not tend  to survive during the CE phase.
But if binary survives in the CE phase, the binary becomes close binary and easy to merge due to the emission of GW.  
Thus, the number of the merging BH-BHs does not decrease.
In 140 case, the number of merging BH-BHs which are low mass becomes larger than our standard model (See Fig. \ref{almass2})  because the low mass BH-BHs can merge more easily than our standard model due to moderately small $\alpha\lambda$.
In Fig. \ref{altime}, the shortest merger time is smaller than that of our standard model due to moderately small $\alpha\lambda$.
$\alpha\lambda$ is small but the binaries which become the CE phase do not tend  to merge during the CE phase and they have close orbit due to small $\alpha\lambda$.

\begin{figure}

  \includegraphics[scale=0.6]{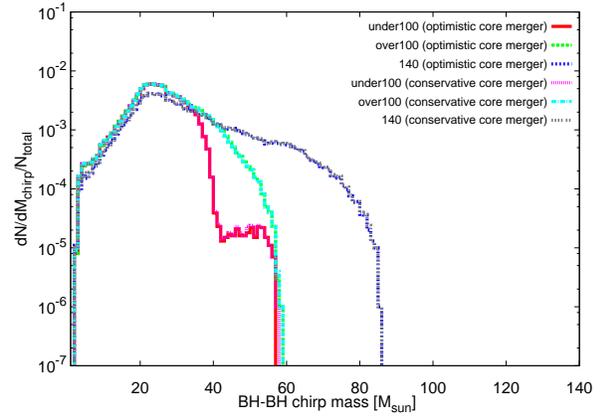} 
\smallskip

\caption{$\alpha\lambda =10$
}

Same as Fig.\ref{our standardmass} but for $\alpha\lambda=10$ model.

\label{almass3}
\end{figure}

Thirdly, we consider the $\alpha\lambda =10$ model. 
In this case, the number of the BH-BH formation increases compared with our standard model.
While  the number of the merging BH-BHs decreases compared with our standard model.
$\alpha\lambda$ is  so large that the separation after the CE phase tends to become large and the binary does not  merge during the CE phase.
Thus, the number of the BH-BH formations increases.
But, due to the large separation  the BH-BH binary does not  merge within 15 Gyrs.
Therefore, the number of the merging BH-BHs decreases.
On the other hand, the number of merging high chirp mass BH-BHs is  more than our standard model (See Fig. \ref{almass3}), because the BH-BHs which are formed after the CE phase tend to have wide orbit due to high $\alpha\lambda$ and the wide massive BH-BHs can merge more easily than the wide low mass BH-BHs.
{Especially, in  140 case this effect is remarkably clear. } 
In Fig. \ref{altime}, the shortest merger time becomes larger than that of our standard model and
the number of merging BH-BHs which merge at the early universe is smaller than that of our standard model due to the same reason.

\subsubsection*{(f)$\beta=0$, $\beta=0.5$ and $\beta=1$ models}
In these cases, the accretion rate during the RLOF is changed by the loss fraction of transferred stellar mass $\beta$.
Firstly, we consider the $\beta=0$ model.

\begin{figure}

  \includegraphics[scale=0.6]{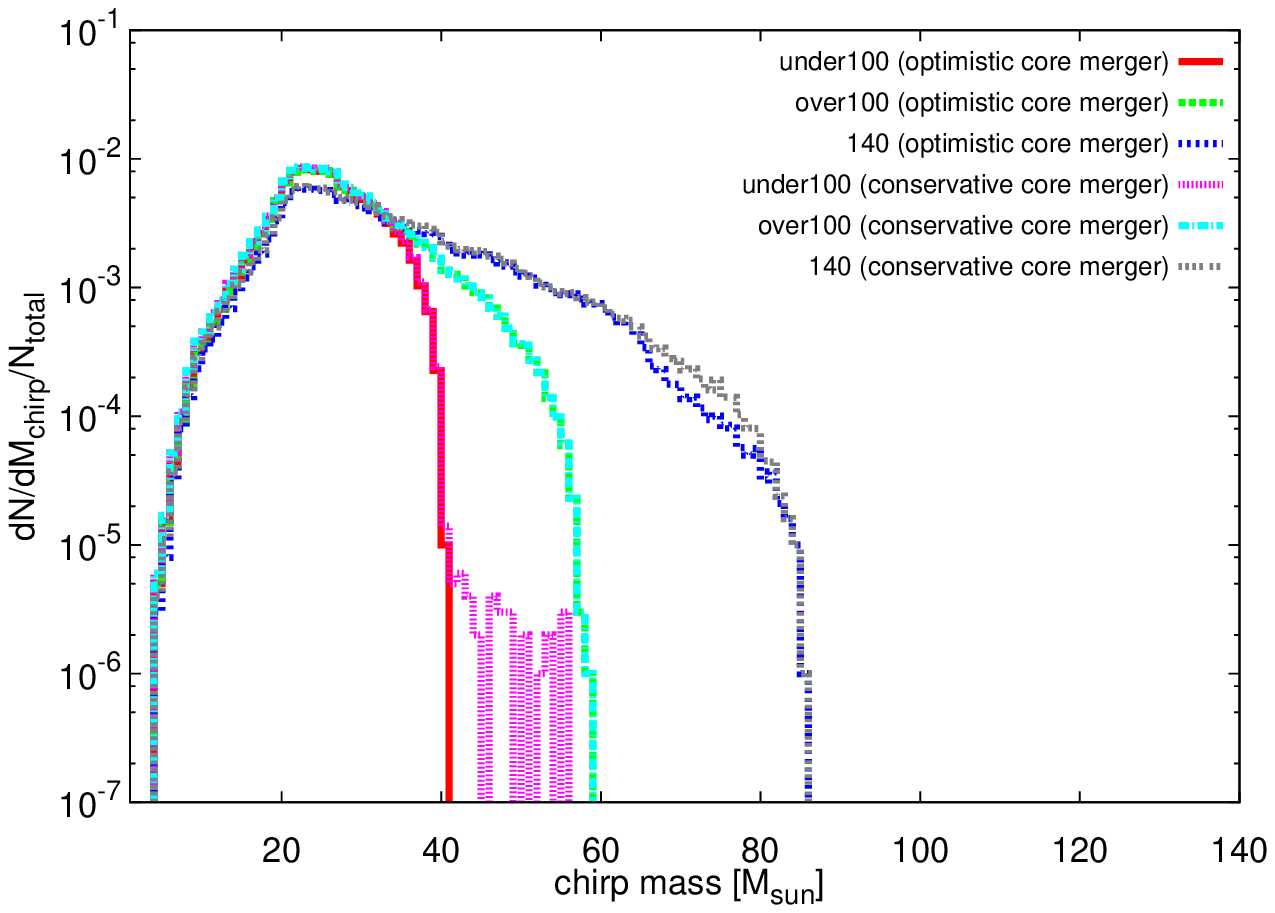} 
\smallskip

\caption{$\beta =0$
}

Same as Fig.\ref{our standardmass} but for $\beta=0$ model.

\label{betamass1}
\end{figure}

The result of our standard model and the result of $\beta=0$ model are the same (Table \ref{model11},\ref{model1} and Figs. \ref{our standardmass}, \ref{betamass1}, \ref{betatime}).
We use the Hurley's fitting $\beta$ which is fitted by Pop I binaries for the Pop III case.
We found the Hurley's fitting $\beta$ is same as $\beta=0$ in Pop III case. 
Fig.\ref{betamass1} and Fig.\ref{our standardmass} are the same.
This means that the RLOF of our standard model (equation \ref{MT}) is the conservative mass transfer prescription.

\begin{figure}
  \includegraphics[scale=0.6]{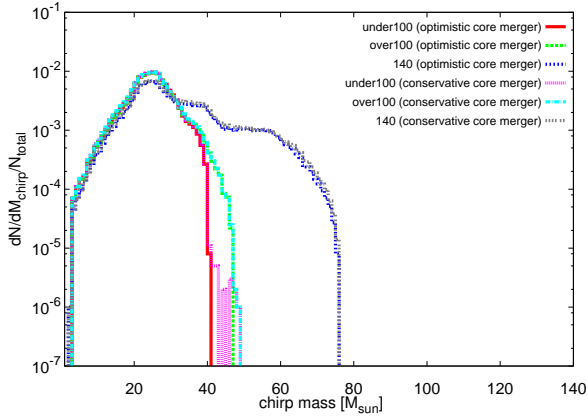} 
\smallskip

\caption{$\beta =0.5$
}

Same as Fig.\ref{our standardmass} but for $\beta=0.5$ model.

\label{betamass2}
\end{figure}

Secondly, we consider $\beta=0.5$ model.
In this case, the mass loss during the mass transfer makes the separation  wide so that the mass transfer tends to be dynamically stable.
The number of the binaries which are merged during the CE phase decreases.
Therefore, the number of BH-BH formation increases compared to our standard model.
However, the number of merging BH-BHs decreases because the binary does not tend to be a close binary by the CE phase.
The peaks of the chirp mass distributions are almost the same as in our standard model (See Fig. \ref{betamass2}).  
 While the highest mass of merging BH-BHs decreases compared with  our standard model because of the mass loss during the RLOF.

\begin{figure}
  \includegraphics[scale=0.6]{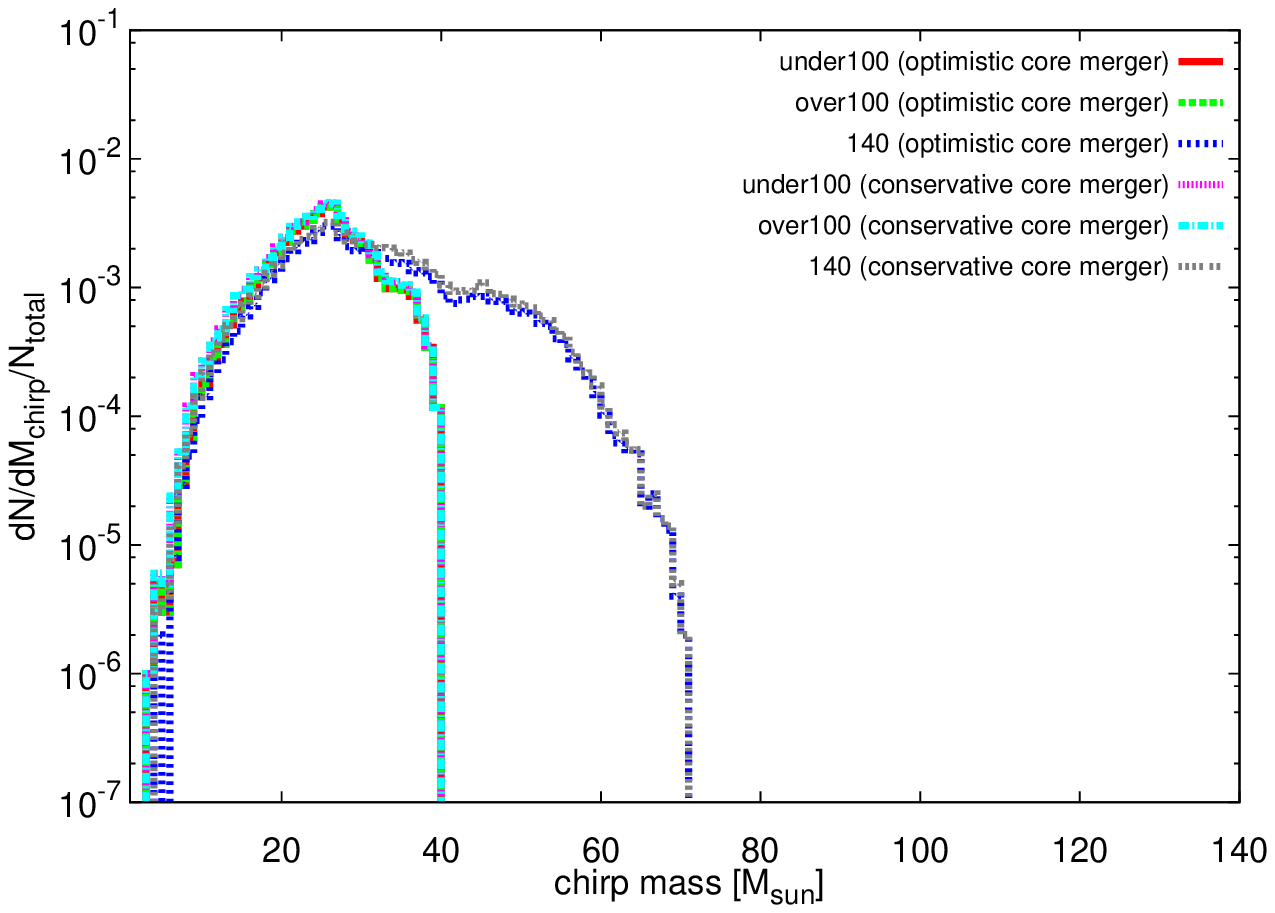} 
\smallskip

\caption{$\beta =1$
}

Same as Fig.\ref{our standardmass} but for $\beta=1$ model.

\label{betamass3}
\end{figure}

Thirdly, we consider  $\beta=1$ model.
In this model, the number of BH-BH formation is larger than   our standard model, but it is smaller than that of $\beta=0.5$ model.
Like  $\beta=0.5$ model, the mass transfer tends to be dynamically stable.
Especially, in this case, there are no paths to the CE phase via dynamically unstable mass transfer.
{There are  only paths to the CE phase in which the secondary plunges into the primary due to the eccentric orbit or in which the each star is a giant and the binary become the contact binary due to the expanding.}
Thus, the number of the binaries which are merged during the CE phase decreases, so that the number of BH-BH formations increases compared with our standard model.
However, the mass loss during the mass transfer is so large that some stars which can be BH by the mass accretion during the RLOF cannot be a BH but a  NS.
Therefore, the  the number of BH-BH formations decreases compared with  $\beta=0.5$ model, but the number of NS-BH formations increases accordingly.
{Furthermore, since the progenitors of merging BH-BHs are hard to enter the CE phase and tend to become wide orbits due to the mass loss during the mass transfer.
Thus, the number of merging BH-BHs decrease.
The major progenitors of merging BH-BHs do not enter the CE phase and lose their mass during the mass transfer. But,  since the Pop III statr radius is small (See Fig.2 in \cite{Marigo2001} and Fig.1 in \cite{Kinugawa2014}), the mass loss during the RLOF tends to stop right away and the separation tends to be  close enough  that Pop III BH-BHs can merge within 15 Gyrs.}
The highest mass peak region of the chirp mass distributions becomes smaller than our standard model and the highest mass of merging BH-BHs decreases due to the mass loss during the RLOF (See Fig. \ref{betamass3}).

\subsubsection*{(g)Worst model}
\begin{figure}

  \includegraphics[scale=0.6]{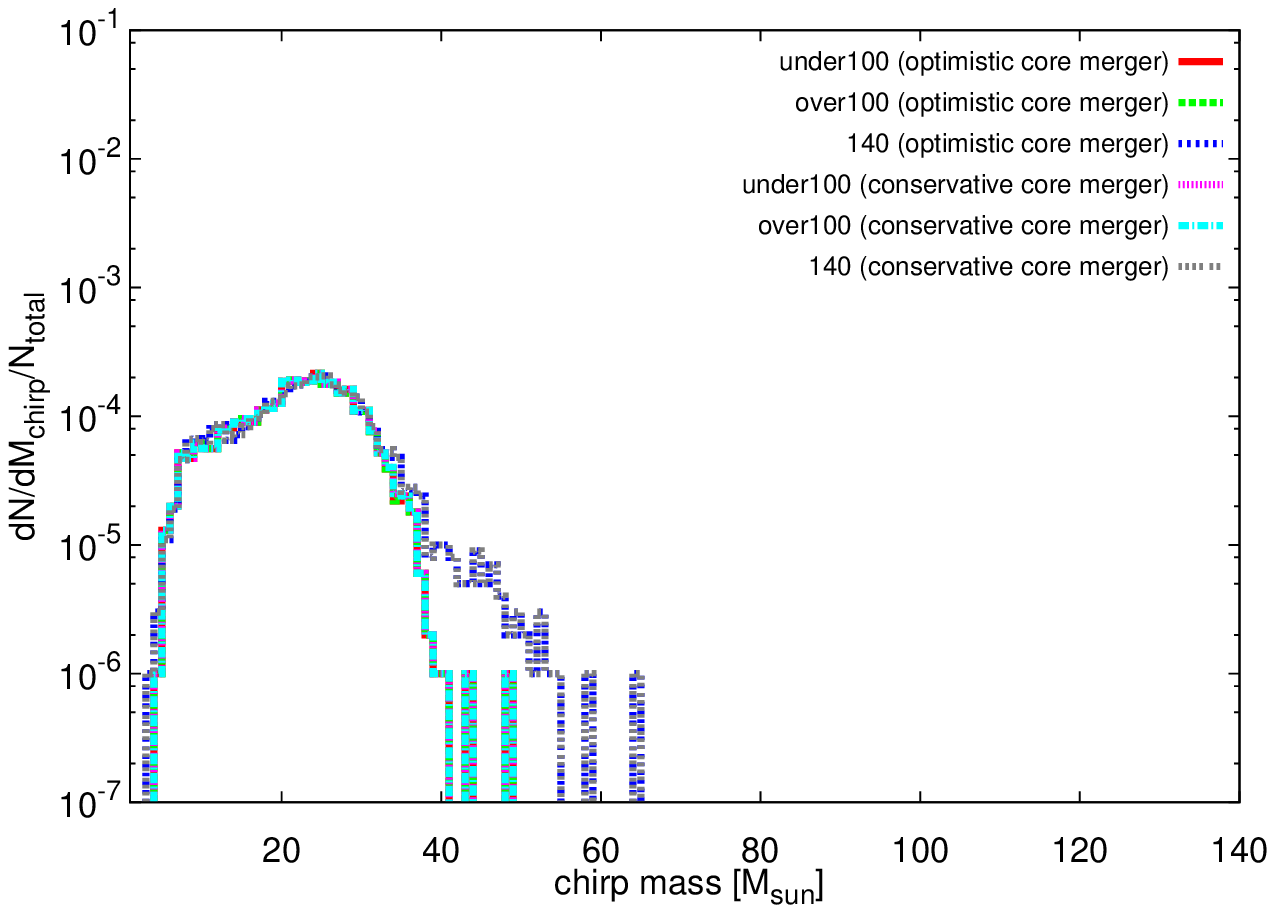} 
\smallskip

\caption{Worst
}

Same as Fig.\ref{our standardmass} but for Worst model.

\label{worstmass}
\end{figure}

\begin{figure}

  \includegraphics[scale=0.6]{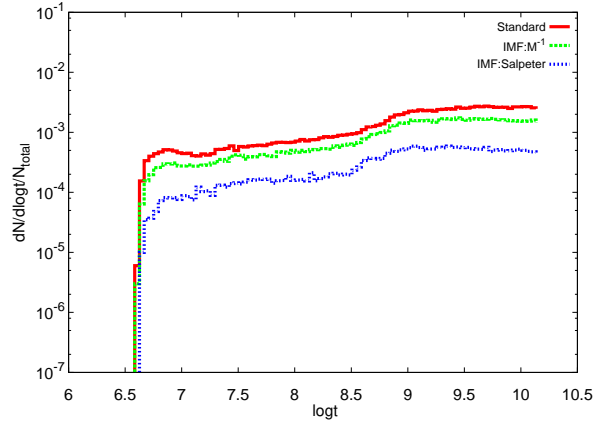} 
\smallskip

\caption{merger time:IMF
}

This figure describes the merger time distributions of Pop III BH-BHs. The red line, the green line and the blue line are our standard model, the logflat model and the Salpeter model. $\rm N_{total}=10^6$ binaries.

\label{IMFtime}
\end{figure}

\begin{figure}

  \includegraphics[scale=0.6]{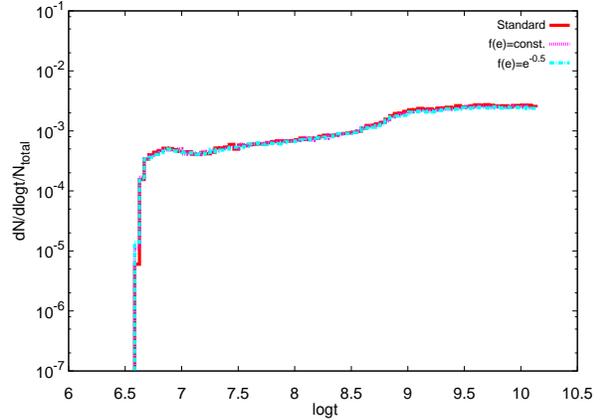} 
\smallskip

\caption{merger time:e
}

This figure describes the merger time distributions of Pop III BH-BHs. The red line, the pink line and the light blue line are our standard model, IEF:const. model and IEF:$e^{-0.5}$ model. $\rm N_{total}=10^6$ binaries.

\label{ecctime}
\end{figure}

\begin{figure}
  \includegraphics[scale=0.6]{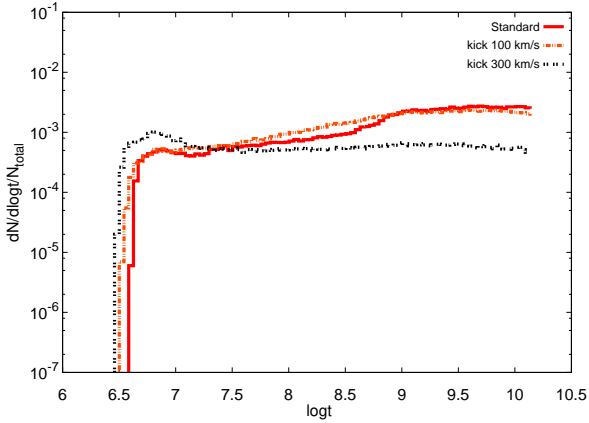} 
\smallskip

\caption{merger time:kick
}

This figure describes the merger time distributions of Pop III BH-BHs. The red line, the orange line and the black line are our standard model, the kick 100 km $s^{-1}$ model and the kick 300 km $s^{-1}$ model. $\rm N_{total}=10^6$ binaries.

\label{kicktime}
\end{figure}

\begin{figure}

  \includegraphics[scale=0.6]{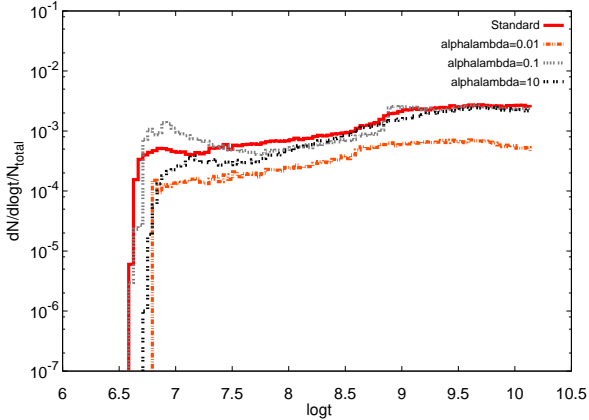} 
\smallskip

\caption{merger time:$\alpha\lambda$
}

This figure describes the merger time distributions of Pop III BH-BHs. The red line, the orange line, the grey line and the black line are our standard model, the $\alpha\lambda=0.01$ model, the $\alpha\lambda=0.1$ and the $\alpha\lambda=10$ model. $\rm N_{total}=10^6$ binaries.

\label{altime}
\end{figure}

\begin{figure}

  \includegraphics[scale=0.6]{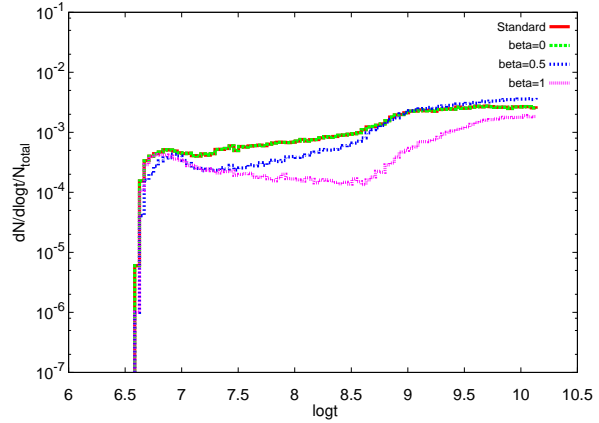} 
\smallskip

\caption{merger time:$\beta$
}

This figure describes the merger time distributions of Pop III BH-BHs. The red line, the green line, the blue line and the pink line are our standard model, the $\beta=0$ model, the $\beta=0.5$ and the $\beta=1$ model. $\rm N_{total}=10^6$ binaries.

\label{betatime}
\end{figure}

\begin{figure}

  \includegraphics[scale=0.6]{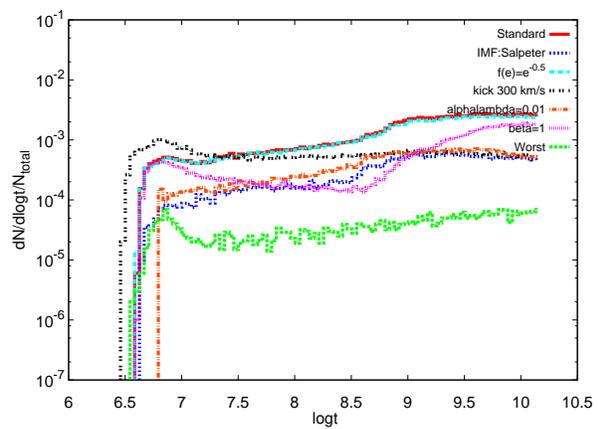} 
\smallskip

\caption{merger time:Worst
}

This figure describes the merger time distributions of Pop III BH-BHs. The red line, the blue line, the light blue line, the black line, the orange line, the pink line and the green line are our standard model, the Salpeter model, the IEF:$e^{-0.5}$ model, the $\alpha\lambda=0.01$ model,  the $\beta=1$ model and the Worst model. $\rm N_{total}=10^6$ binaries.

\label{saiakutime}
\end{figure}

In this model, we choose the initial conditions and binary parameters which will make the worst result in (b)IMF, (c)IEF, (d)kick, (e)$\alpha\lambda$ and (f)$\beta$.
Thus, we adopt (b)IMF:Salpeter, (c)IEF:$e^{-0.5}$, (d)kick 300 $\rm km~s^{-1}$, (e)$\alpha\lambda=0.01$ and (f)$\beta=1$.
Especially, we already know that (b)IMF, (d)kick, (e)$\alpha\lambda$ and (f)$\beta$ influence the result very much so that  the result of BH-BH formation and the number of 
merging BH-BHs are determined by these parameters and IMF.
Each effect of (b)IMF:Salpeter, (d)kick 300 $\rm km~s^{-1}$, (e)$\alpha\lambda=0.01$ and (f)$\beta=1$ makes the number of merging BH-BHs decrease (See Table \ref{model3}, \ref{model7} and \ref{model13}).
Thus, the number of merging BH-BHs extremely decreases compared with our standard model (See Tabel \ref{model1} and \ref{model14}). 
On the properties of chirp mass distribution, the number of merging BH-BHs more massive than $30~\msun$ decreases and the gradient of the chirp mass distribution of merging BH-BHs is much steeper than that of our standard model (See Fig. \ref{worstmass} (a) (b)). 
The Salpeter IMF makes the number of high mass stars to decrease and non-conservative mass transfer ($\beta=1$) prevents to merge BH-BHs while in our standard mode they  can be merging BH-BHs after the CE phase.
Furthermore, even though the massive binaries become the CE phase, they usually merge during the CE phase due to the very small $\alpha\lambda$. 
Therefore, the peak region of the chirp mass is 25-30 $\msun$ even in 140 case.

\begin{table*}
\caption{our standard model
}

{This table shows  the numbers of NS-NS, NS-BH and BH-BH binaries and the numbers of each compact binary which merges within 15 Gyrs for our standard model. 
 15Gyrs is used in order to compare our results with previous works. To estimate the present merger rates, we use 13.8Gyrs as the present age of the universe. The meanings of  under 100, over 100 and 140 are explained 
 in the first paragraph of \S 2.  
The numbers in the parenthesis  are for the case of the 
conservative core-merger criterion while those without the parenthesis are for the case of the optimistic 
core-merger criterion. The meanings of these two criteria were explained in the last paragraph of \S 2 before \S 2.1}

\label{model1}
\begin{center}
\begin{tabular}{c c c c} 
\hline
 & under100 & over100 &140\\ 
\hline
NS-NS& 0 (279) & 0 (279) & 0 (195)\\
NS-BH& 185335 (187638) & 185335 (187638) & 153435 (155694) \\
BH-BH& 517067 (522581) & 534693 (540316) & 595894 (604930) \\
merging NS-NS& 0 (279) & 0 (279) & 0 (195)\\ 
merging NS-BH& 50 (149) & 50 (149) & 825 (1255) \\ 
merging BH-BH& 115056 (120532) & 131060 (136645) & 128894 (137903)\\  \hline
\end{tabular}
\end{center}
\end{table*}
\begin{table*}
\caption{IMF:logflat
}
\label{model2}

{Same as Table \ref{model1} but for IMF:logflat model.}

\begin{center}
\begin{tabular}{c c c c} 
\hline
 & under100 & over100 &140\\ 
\hline
NS-NS& 2 (789) & 2 (789) & 1 (693)\\
NS-BH& 168100 (169794) & 168100 (169794) & 157106 (158831) \\
BH-BH& 350169 (353524) & 357989 (361378) & 405922 (410802) \\
merging NS-NS& 2 (789) & 2 (789) & 1 (693)\\
merging NS-BH& 68 (183) & 68 (183) & 374 (579) \\ 
merging BH-BH& 74745 (78054) & 81786 (85129) & 87590 (92450) \\  \hline
\end{tabular}
\end{center}
\end{table*}

\begin{table*}
\caption{IMF:Salpeter
}
\label{model3}

{Same as Table \ref{model1} but for IMF:Salpeter model.}

\begin{center}
\begin{tabular}{c c c c} 
\hline
 & under100 & over100 &140\\ 
\hline
NS-NS& 5 (1994) & 5 (1994) & 3 (1957)\\
NS-BH& 93085 (93793) & 93085 (93793) & 92861 (93603) \\
BH-BH& 132534 (133485) & 133880 (134835) & 144096 (145294) \\
merging NS-NS& 5 (1994) & 5 (1994) & 3 (1957)\\
merging NS-BH& 64 (164) & 64 (164) & 97 (216) \\ 
merging BH-BH& 25536 (26468) & 26720 (27656) & 28378 (29564) \\  \hline
\end{tabular}
\end{center}
\end{table*}

\begin{table*}
\caption{IEF:const.}
\label{model4}

{Same as Table \ref{model1} but for IMF:logflat model.}

\begin{center}
\begin{tabular}{c c c c} 
\hline
 & under100 & over100 &140\\ 
\hline
NS-NS& 0 (358) & 0 (358) & 0 (255)\\
NS-BH& 183460 (184761) & 183460 (184761) & 152099 (153548) \\
BH-BH& 522809 (526892)  & 541264 (545459) & 602071 (608210) \\
merging NS-NS& 0 (358) & 0 (358) & 0 (255)\\
merging NS-BH& 43 (130) & 43 (130) & 843 (1087) \\ 
merging BH-BH& 111106 (1115171) & 127904 (132081) & 124714 (130831) \\  \hline
\end{tabular}
\end{center}
\end{table*}

\begin{table*}
\caption{IEF:$e^{-0.5}$
}
\label{model5}

{Same as Table \ref{model1} but for IEF:$e^{-0.5}$ model.}

\begin{center}
\begin{tabular}{c c c c} 
\hline
 & under100 & over100 &140\\ 
\hline
NS-NS& 0 (365) & 0 (365) & 0 (258)\\
NS-BH& 181650 (182388) & 181650 (182388) & 150779 (151805) \\
BH-BH& 523285 (526534) & 542015 (545389) &602575 (607054) \\
merging NS-NS& 0 (365) & 0 (365) & 0 (258)\\
merging NS-BH& 38 (100) & 38 (100) & 774 (964) \\ 
merging BH-BH& 107594 (110832) & 124620 (127983) & 121494 (125955)\\  \hline
\end{tabular}
\end{center}
\end{table*}

\begin{table*}
\caption{kick 100 $\rm km~s^{-1}$
}
\label{model6}

{Same as Table \ref{model1} but for kick 100 $\rm km~s^{-1}$ model.}

\begin{center}
\begin{tabular}{c c c c} 
\hline
 & under100 & over100 &140\\ 
\hline
NS-NS& 283 (794) & 283 (794) & 180 (516)\\
NS-BH& 32701 (34778) & 32701 (34778) & 32014 (34144) \\
BH-BH& 191755 (197327) & 208268 (213962) &234117 (243348) \\
merging NS-NS& 17 (526) & 17 (526) & 6 (342)\\
merging NS-BH& 2527 (3016) & 2527 (3016) & 3218 (3762) \\ 
merging BH-BH& 117415 (122830) & 132066 (137603) & 135758 (144554)\\  \hline
\end{tabular}
\end{center}
\end{table*}

\begin{table*}
\caption{kick 300 $\rm km~s^{-1}$
}
\label{model7}

{Same as Table \ref{model1} but for kick 300 $\rm km~s^{-1}$ model.}

\begin{center}
\begin{tabular}{c c c c} 
\hline
 & under100 & over100 &140\\ 
\hline
NS-NS& 8 (112) & 8 (112) & 4 (78)\\
NS-BH& 11922 (13133) & 11941 (13152) & 12115 (13330) \\
BH-BH& 70728 (75011) & 78058 (82496) & 86876 (93481) \\
merging NS-NS& 1 (85) & 1 (85) & 1 (60)\\
merging NS-BH& 3893 (4483) & 3900 (4490) & 4406 (5002) \\ 
merging BH-BH& 51928 (56021) & 58793 (63041) & 64084 (70252)\\  \hline
\end{tabular}
\end{center}
\end{table*}

\begin{table*}
\caption{$\alpha\lambda = 0.01$
}
\label{model8}

{Same as Table \ref{model1} but for $\alpha\lambda=0.01$ model.}

\begin{center}
\begin{tabular}{c c c c} 
\hline
 & under100 & over100 &140\\ 
\hline
NS-NS& 0 (0) & 0 (0) & 0 (0)\\
NS-BH& 148290 (148770) & 148290 (148770) & 116548 (117117) \\
BH-BH& 340893 (352047) & 345140 (363191) & 365526 (382686) \\
merging NS-NS& 0 (0) & 0 (0) & 0 (0)\\
merging NS-BH& 0 (294) & 0 (294) & 30 (412) \\ 
merging BH-BH& 32283 (43437) & 36530 (54581) & 27790 (44950)\\  \hline
\end{tabular}
\end{center}
\end{table*}

\begin{table*}
\caption{$\alpha\lambda = 0.1$
}
\label{model9}

{Same as Table \ref{model1} but for $\alpha\lambda=0.1$ model.}

\begin{center}
\begin{tabular}{c c c c} 
\hline
 & under100 & over100 &140\\ 
\hline
NS-NS& 0 (0) & 0 (0) & 0 (0)\\
NS-BH& 162814 (173016) & 162814 (173016) & 130556 (138835) \\
BH-BH& 434590 (464369) & 448847 (480217) & 480520 (520031) \\
merging NS-NS& 0 (0) & 0 (0) & 0 (0)\\
merging NS-BH& 45 (181) & 45 (181) & 1065 (1877) \\ 
merging BH-BH& 111696 (141356) & 125953 (157204) & 124830 (164240)\\  \hline
\end{tabular}
\end{center}
\end{table*}

\begin{table*}
\caption{$\alpha\lambda = 10$
}
\label{model10}

{Same as Table \ref{model1} but for $\alpha\lambda=10$ model.}

\begin{center}
\begin{tabular}{c c c c} 
\hline
 & under100 & over100 &140\\ 
\hline
NS-NS& 1116 (2215) & 1116 (2215) & 840 (1616)\\
NS-BH& 198408 (198758) & 198408 (198758) & 166173 (166408) \\
BH-BH& 542399 (542603) & 560156 (560360) & 624631 (624958) \\
merging NS-NS& 890 (1949) & 890 (1949) & 634 (1381)\\
merging NS-BH& 767 (975) & 767 (975) & 506 (645) \\ 
merging BH-BH& 91787 (91989) & 104656 (104858) & 93729 (94055)\\  \hline
\end{tabular}
\end{center}
\end{table*}

\begin{table*}
\caption{$\beta = 0$
}
\label{model11}

{Same as Table \ref{model1} but for $\beta=0$ model.}

\begin{center}
\begin{tabular}{c c c c} 
\hline
 & under100 & over100 &140\\ 
\hline
NS-NS& 0 (279) & 0 (279) & 0 (195)\\
NS-BH& 185335 (187638) & 185335 (187638) & 153435 (155694) \\
BH-BH& 517067 (522581) & 534693 (540316) & 595894 (604930) \\
merging NS-NS& 0 (279) & 0 (279) & 0 (195)\\ 
merging NS-BH& 50 (149) & 50 (149) & 825 (1255) \\ 
merging BH-BH& 115056 (120532) & 131060 (136645) & 128894 (137903)\\  \hline
\end{tabular}
\end{center}
\end{table*}

\begin{table*}
\caption{$\beta = 0.5$
}
\label{model12}

{Same as Table \ref{model1} but for $\beta=0.5$ model.}

\begin{center}
\begin{tabular}{c c c c} 
\hline
 & under100 & over100 &140\\ 
\hline
NS-NS& 5 (380) & 5 (380) & 6 (272)\\
NS-BH& 193921 (196094) & 193921 (196094) & 158518 (160442) \\
BH-BH& 549893 (554150) & 554966 (559228) & 628253 (635698) \\
merging NS-NS& 5 (380) & 5 (380) & 6 (272)\\
merging NS-BH&  199 (286) & 199 (286) & 766 (1082) \\ 
merging BH-BH& 117094 (121310) & 119758 (123979) & 126090 (133512)\\  \hline
\end{tabular}
\end{center}
\end{table*}

\begin{table*}
\caption{$\beta = 1$
}
\label{model13}

{Same as Table \ref{model1} but for $\beta=1$ model.}

\begin{center}
\begin{tabular}{c c c c} 
\hline
 & under100 & over100 &140\\ 
\hline
NS-NS& 1359 (2006) & 1359 (2006) & 898 (1344)\\
NS-BH& 218311 (220521) & 218311 (220522) & 178444 (180375) \\
BH-BH& 531452 (536579) & 531484 (536611) & 610732 (619230) \\
merging NS-NS& 1358 (2005) & 1358 (2005) & 898 (1344)\\
merging NS-BH& 119 (255) & 119 (255) & 578 (917) \\ 
merging BH-BH& 50119 (55214) & 50119 (55214) & 57025 (65121)\\  \hline
\end{tabular}
\end{center}
\end{table*}

\begin{table*}
\caption{Worst
}
\label{model14}

{Same as Table \ref{model1} but for Worst model.}

\begin{center}
\begin{tabular}{c c c c} 
\hline
 & under100 & over100 &140\\ 
\hline
NS-NS& 1637 (1637) & 1637 (1637) & 1604 (1604)\\
NS-BH& 4345 (4345) & 4345 (4345 & 4283 (4285) \\
BH-BH& 5227 (5235) & 5227 (5235) & 5560 (5586) \\
merging NS-NS& 1562 (1562) & 1562 (1562) & 1532 (1532)\\
merging NS-BH& 1645 (1645) & 1645 (1645) & 1604 (1606) \\ 
merging BH-BH& 3195 (3203) & 3195 (3203) & 3376 (3399)\\  \hline
\end{tabular}
\end{center}
\end{table*}

\clearpage



\clearpage

\subsection{The merger rate properties of Pop III BH-BHs}
\begin{figure*}
\centering
\includegraphics[width=12cm]{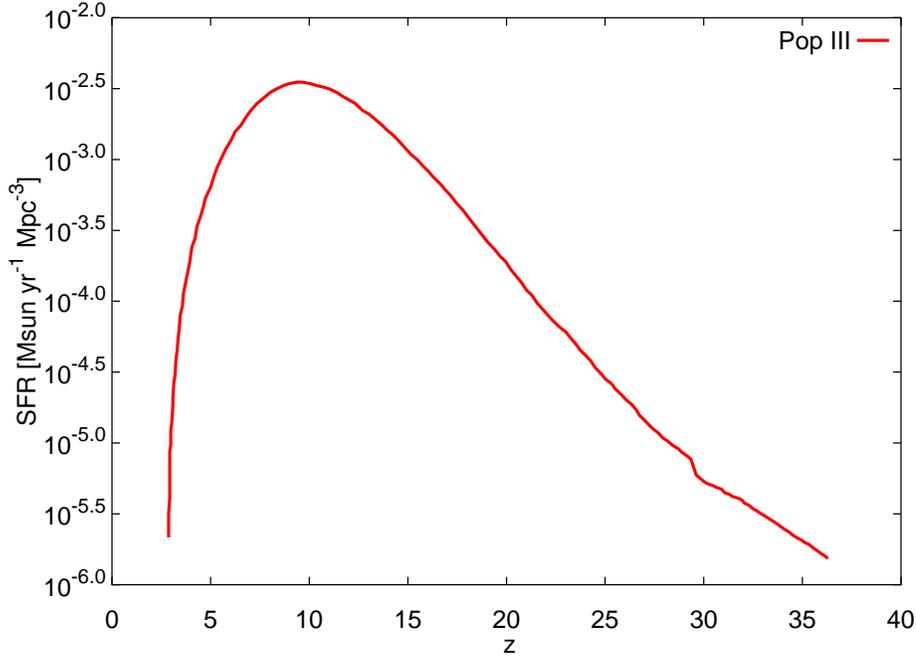}
\caption{The star formation rate density (comoving) calculated by de Souza et al. (2011). The unit of the rate is $\msun$ per comoving volume per proper time. The red line is the the total SFR density of Pop III stars.}
\label{sfr}
\end{figure*}

In this sub-section, we show the merger rate density of each model.
The merger rate density $R_{i}(t)$ is calculated using results of each model and the Pop III star formation rate of \cite*{de Souza2011} ({See Fig. \ref{sfr}} and \cite{Kinugawa2014} for details) as
\begin{equation}
R_{i}(t)=\int^t_0\frac{\rm f_b}{\rm 1+f_b}\frac{{\rm SFR}(t')}{<M>}\frac{N_i(t-t')}{N_{\rm total}}dt'
\end{equation}
where $i$ is the type of compact binaries such as NS-NS, NS-BH or BH-BH.
$f_b$ is the initial binary faction.
The recent cosmological hydrodynamics simulation \citep{Susa2014} suggests that the binary fraction is about 50\%.
Thus, we use $\rm f_b=1/2$ since the total number of the binary is half of the total number of the stars in the binary.
 $<M>$ is the mean initial stellar mass of Pop III star derived from IMF and the initial mass range.
${\rm{SFR}}(t')~[\msun\rm ~yr^{-1}~Mpc^{-3}]$ is the Pop III star formation rate at $t'$.
$N_i(t-t')$ is the number of type $i$ compact binaries which are formed in $[t',t'+dt]$ and merge at  time $t$.
$N_{\rm total}$ is the total number of the simulated binaries.

\begin{figure}
\medskip
  \includegraphics[scale=0.6]{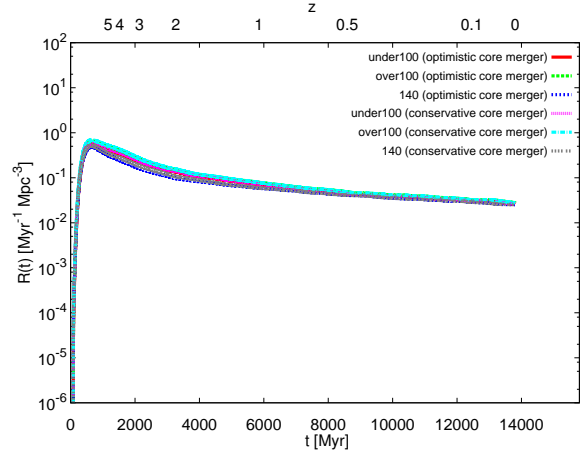}

\caption{our standard
}

The red, green, blue, pink, light blue and grey  lines are the under100 case with optimistic core-merger criterion, the over100 case with optimistic core-merger criterion, the 140 case with optimistic core-merger criterion, the under100 case with conservative core-merger criterion, the over100 case with conservative core-merger criterion and the 140 case with conservative core-merger criterion, respectively.

\label{our standardrate}
\end{figure}

Fig. \ref{our standardrate} and not show the merger rate densities $\rm [Myr^{-1}~Mpc^{-3}]$ of  BH-BHs as a function of cosmic time (lower abscissa) and redshift $z$ (upper abscissa) in our standard model and the worst model.
It is seen that in each model the  merger rate densities for the same redshift  depend on neither the  initial mass range ( [10,100] or [10,140] ) nor the CE merger criterion.
The other models have the same dependencies so that we do not show their figures.
As a function of the redshift, the merger rate densities are nearly constant from $z=0$ to  $z\sim1$ in each model.
Table \ref{ratez=0} shows the  merger rate density $\rm [Myr^{-1}~Mpc^{-3}]$ at $z=0~{(t_{\rm Hubble}=13.8~\rm Gyrs)}$ for each model.
The lowest rate is as expected  in worst model while the highest rate is in $\beta=0.5$ model.

Fig. \ref{mergerz} shows  the difference of the merger rate density of each model for  under100 case.
Table \ref{peak} describes the peak redshift of the BH-BHs merger rate density of each model in  under100 case. 
It is seen that the peak redshift of the BH-BHs merger rate density ranges from 8.8 to 7.15.
These peak redshifts are   near the peak of the star formation rate at $z\sim 9$.
In the following, we discuss the difference of each model.

\begin{figure}
\medskip
  \includegraphics[scale=0.6]{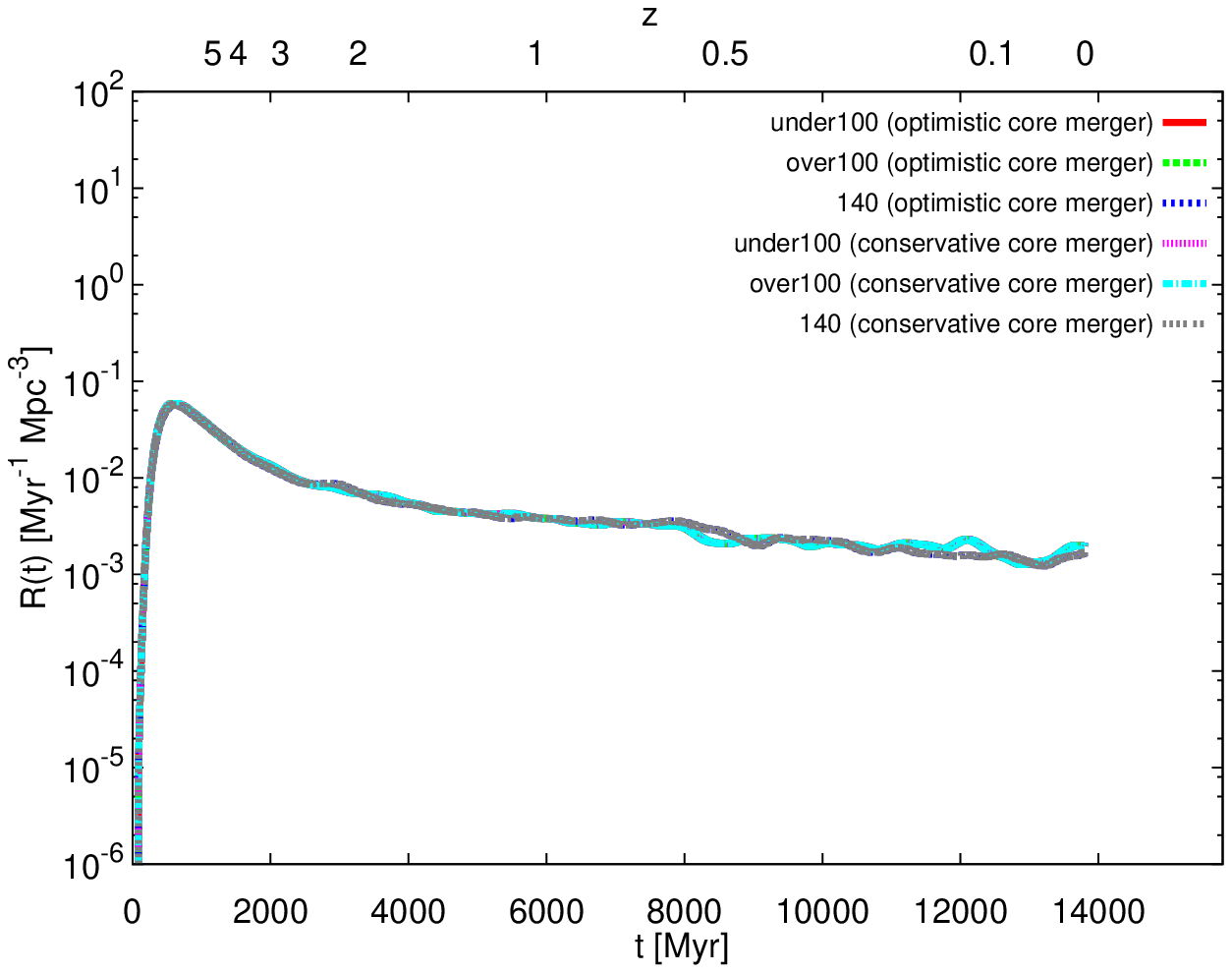}

\caption{Worst
}

The same as Fig. \ref{our standardrate} but for Worst model.

\label{worstrate}
\end{figure}

\begin{table*}
\caption{The merger rate density [$\rm Myr^{-1}~Mpc^{-3}$] at $z=0 (t_{\rm Hubble}=13.8~\rm Gyrs)$
}
\label{ratez=0}
\begin{center}
\begin{tabular}{c c c c} 
\hline
                 &        under100        &        over100              &            140         \\
\hline
our standard       &  0.0258 (0.0260) & 0.0277 (0.0279) & 0.0251 (0.0252)\\
IMF:logflat  &   0.0230 (0.0232)& 0.0240 (0.0245) & 0.0232 (0.0236)\\
IMF:Salpeter & 0.0116 (0.0117) & 0.0121 (0.0122) & 0.0131 (0.0132)\\
IEF:const.  & 0.0267 (0.0267) & 0.0288 (0.0288) & 0.0242 (0.0242)\\
IEF:$e^{-0.5}$  & 0.0252 (0.0252) & 0.0270 (0.0271) & 0.0228 (0.0228)\\
kick 100 $\rm km~s^{-1}$ &    0.0210 (0.0212) & 0.0223 (0.0226) & 0.0203 (0.0207)\\
kick 300 $\rm km~s^{-1}$ &    0.00726 (0.00732) & 0.00747 (0.00754) & 0.00657 (0.00672)\\
$\alpha\lambda=0.01$    &   0.00542 (0.00542) & 0.00542 (0.00542) & 0.00290 (0.00290)\\
$\alpha\lambda=0.1$     &   0.0249 (0.0255) & 0.0249 (0.0255) & 0.0207 (0.0210)\\
$\alpha\lambda=10$      &   0.0229 (0.0229) & 0.0253 (0.0253) & 0.0192 (0.0192)\\
$\beta=0$    &   0.0258 (0.0260) & 0.0277 (0.0279) & 0.0251 (0.0252)\\
$\beta=0.5$ &   0.0361 (0.0362) &0.0369 (0.0370) & 0.0320 (0.0321)\\
$\beta=1$ & 0.0186 (0.0187) & 0.0186 (0.0187) &0.0159 (0.0161)\\
Worst& 0.00194 (0.00194) & 0.00194 (0.00194) & 0.00169 (0.00169)\\
\hline
\end{tabular}
\end{center}
\end{table*}

The IMF dependence of the peak redshift of the merger rate density is clear seen. Namely
 for the  steeper IMF  the peak redshift   is small although the difference is not so large ($\sim$0.45 in $z$).
Since  BH-BH progenitors whose initial mass is lower than 50 $\msun$ tend to evolve via the RLOF but not via the CE, the steeper IMF  can  make BH-BH progenitors to evolve via RLOFs. 
BH-BHs which evolved via RLOFs tend to have the wider orbit than BH-BHs which evolved via CE phases.
Therefore, the typical merger time for the steeper IMF tends to be long so that the peak redshift is smaller. 
As for  the IEF dependence, no tendency is seen while
 as for the natal kick velocity dependence, the peak redshift for 100 $\rm km~s^{-1}$ model is smaller than our standard model, but that of the  300 $\rm km~s^{-1}$ model is higher than our standard model. In the  100 $\rm km~s^{-1}$ model, the kick makes the BH-BHs to eccentric orbit so that the merger time becomes smaller than that of the circular orbit.
Thus, the number of the merging BH-BHs which merge at the high redshift should increase.
However, the BH-BHs which cannot merge in our standard case due to wide orbit can merge due to the natal kick.
Consequently, the number of the merging BH-BHs which merge at lower redshift tends to increase.
In the  300 $\rm km~s^{-1}$ model, the natal kick velocity is too large so that the binary tends to disrupt.
However, if the natal kick direction is against the orbital direction, the natal kick behaves like the brake of the car or if the separation before the natal kick is very close, the binary can survive.
Thus, the survived binary tends to have very close and eccentric orbit so that  they can merge early. This explains the apparent strange behavior of the dependence of the 
peak redshift on the natal kick velocity.
 
In the case of the CE parameter, it changes the number of survived binaries during the CE phase and the merger time of the BH-BHs.
If  $\alpha\lambda$ is low, i.e. the orbit energy loss via the CE phase is high, the number of survived binaries during the CE phase is  small and the merger time of the BH-BHs is short. In $\alpha\lambda=0.01$ model, the parameter $\alpha\lambda$ is so small that the almost all binaries which enter the CE phase merge during the CE phase.
Thus, merging BH-BH progenitors evolved via RLOF so that  they have wide orbit.
Therefore, their merger time tend to be long and the peak redshift is smaller  than our standard model.
In $\alpha\lambda=0.1$ model, $\alpha\lambda$ is small but the binaries which become the CE phase do not  merge during the CE phase and they have close orbit due to small $\alpha\lambda$. Thus, the merger time of BH-BHs is  short and the peak redshift is large.     
In $\alpha\lambda=10$ model, $\alpha\lambda$ is so large that  binaries after the CE phase have wide orbit due to large $\alpha\lambda$.
Thus, the merger time of BH-BHs is  long and the peak redshift is small. These consideration explains the strange behavior of   the peak redshift on
$\alpha\lambda$ parameter.

In the case of the parameter  $\beta$, it not only changes the mass accretion to the secondary but also changes the criterion of the dynamically unstable mass transfer.
In $\beta=0.5$ model, the mass transfer is dynamically stable so that the number of the binaries which evolve via RLOF but not via the CE phase.
Thus, the typical merger time is long and the peak redshift is low.
In $\beta=1$ model, the mass transfer is always dynamically stable.
Furthermore, the mass accretion to the secondary during RLOF does not occur so that  the orbit becomes wide, 
On the other hand, binaries which have the eccentric orbit have  only  the CE phase.
Thus, the merging BH-BHs are separated into  two groups. In one group the binaries evolve via RLOF while in the other group  they  evolve via the CE phase due to the eccentric orbit. The former group merges at low redshift and the latter group does at high redshift.
Therefore, in the $\beta=1$ model, the merger rate density are bimodal as shown in Fig. \ref{mergerz}.

{Note that the maximum merger rate density of $3.7 \times 10^{-8}\ {\rm events}\ 
{\rm yr}^{-1}\ {\rm Mpc}^{-3}$ from Table \ref{ratez=0} with $\beta=0.5$ and over100 case is consistent with
the upper limit of $\sim 10^{-7}\ {\rm events}\ {\rm yr}^{-1}\ {\rm Mpc}^{-3}$ by LIGO-Virgo(S6/VSR2/VSR3)~\citep{Aasi2013}.}

\begin{table} 
\caption{The peak redshift of the BH-BHs merger rate density }
\label{peak}
\begin{center}
\begin{tabular}{c c } 
\hline
 model & peak redshift \\ 
\hline
our standard& 7.85\\
IMF:logflat& 7.75 \\
IMF:Salpeter& 7.4 \\
IEF:const.&7.85\\
IEF:$e^{-0.5}$& 7.8\\ 
kick 100 $\rm km~s^{-1}$& 7.5 \\  
kick 300 $\rm km~s^{-1}$& 8.65\\
$\alpha\lambda=0.01$& 7.2\\
$\alpha\lambda=0.1$& 8.5\\
$\alpha\lambda=10$&6.85\\
$\beta=0$&7.85\\
$\beta=0.5$&7.15\\
$\beta=1$&8.8\\
Worst&  8.3\\\hline
\end{tabular}
\end{center}
\end{table}

\begin{figure}
\medskip
  \includegraphics[scale=0.6]{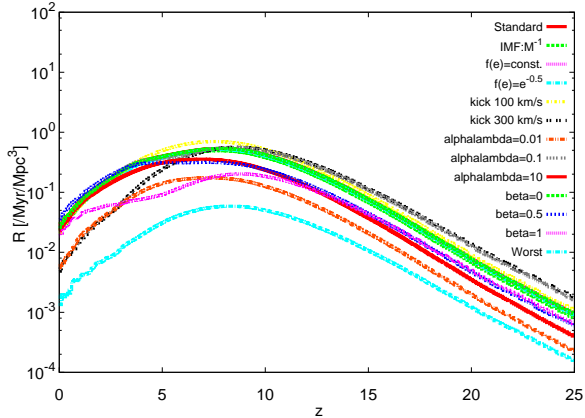}

\caption{The merger rate densities}
\label{mergerz}
\end{figure}

\subsection{The detection rate of Pop III BH-BHs by the second generation detectors}
In this section, we show how to calculate the detection rate of BH-BHs.
Our Pop III population synthesis simulations produced a set of merging BH-BHs with component masses and merger time.
{To estimate the number of the event and the parameter decision accuracy according to
the generated binary mass distribution, distances, various incident angles and
orientations of the orbit plane, we employ the simple Monte-Carlo simulation. As the
typical 2nd generation detectors that has a little advantage in a low frequency band
in underground site, we employ KAGRA detector in our simulation. 
We use the detection range of \cite{Kanda2011}.
The official sensitivity limit of the KAGRA\footnote{http://gwcenter.icrr.u-tokyo.ac.jp/en/researcher/parameter} is suitable for the detection of both inspiral and ringdown gravitational waves from the 10-30 $\msun$ binaries.
In the Monte-Carlo simulation, we placed each event at random position in the hemisphere, random direction of the binary orbit plane.
The direction in the cosmological redshift and the mass are given by the Pop III binary simulation. 
We iterate many events for 1000 years, then, we estimate the expected detection rate for one year observation. The error of the rate is given as the square root of the number of detection in Monte-Carlo trials.}

{Tables \ref{detecthu100} to  \ref{detectb140} describe the detection rate of BH-BHs in each model.
This table shows the detection rates of Pop III BH-BHs for under100 cases with the optimistic core-merger criterion.
The first column
shows the name of the model. The second column shows the detection rate only by the inspiral chirp signal.
The third, the fourth and the fifth columns show the
detection rate only by the quasi normal mode (QNM) with Kerr parameter a/M = 0.70,
the detection rate by the quadrature sum of the inspiral chirp signal and the QNM with a/M = 0.70 and the detection rate by the linear sum of the inspiral chirp signal and the QNM with a/M = 0.70, respectively.
The sixth, the seventh and the eighth columns show the detection rates only by the QNM with a/M = 0.98, the detection rate by the quadrature sum of the inspiral chirp signal and the QNM with a/M = 0.98 and the detection rate by the linear sum of the inspiral chirp signal and the QNM with a/M = 0.98, respectively.
When signal-to-noise ratio of event that is calculated by matched filtering equation, 
over threshold S/N = 8, the event is detected.}
The QNM S/N is calculated by Eq. B14 in \cite*{Flanagan1998}.
$\epsilon_r$ in this equation is the fraction of binary total mass energy radiated in
the QNM.
We assumed the value $\epsilon_r = 0.03$.
Since their equation is averaged over the GW polarization and the sky location, a factor
1/5 is multiplied by the equation. 
However we have to take account of angular values of binary, we replace the factor
with 
$\sqrt{ (1 + \cos^2 \iota)^2 /4 \cdot F_+^2 + \cos^2 \iota \cdot F_\times^2 }$, where the
$\iota$ is 
the inclination angle, and $F_+, F_\times$ are KAGRA antenna pattern functions.
For the fourth and sixth columns, their S/N are calculated by the linear summation
of S/N of the inspiral and the QNM with a/M = 0.70 and 0.98, respectively.
All the rates are based on 1000 years Monte Carlo simulations.

In \cite{Zlochower2015}, the reasonable value of Kerr parameter $a/M$ is about 0.7.
{Thus, we focus on the detection rate by the quadrature sum of the inspiral chirp signal and the QNM with $a/M=0.70$.
The rates of the quadrature sum of the inspiral and the QNM is about 1/2 of the rates of the linear sum of the inspiral and the QNM.
In  our standard model with under100 case and optimistic core-merger criterion, the detection rate by the quadrature sum of the inspiral chirp signal and the QNM is $\sim$1.8 $\times10^2~\rm events~yr^{-1}~(\rm SFR_p/(10^{-2.5}~\msun \rm~yr^{-1}~Mpc^{-3}))\cdot([f_b/(1+f_b)]/0.33)$ where $\rm SFR_{p}$ and $\rm f_b$ are the peak value of the Pop III star formation rate and the binary fraction.
$\rm Err_{sys}=1$ corresponds to the rate  for our standard model with under100 case and the optimistic core-merger criterion.
The definition of $\rm Err_{sys}$ is slightly different from that in our previous paper \citep{Kinugawa2014}. 
That is, the new definition is based on Monte Carlo simulations of the detection of  the inspiral chirp signal and QNM with $a/M=0.7$. 
The basic numerical data of population synthesis of our standard model is the same.
Then Tables \ref{detecthu100}-\ref{detectb140} show that $\rm Err_{sys}$ ranges from  $4.6\times10^{-2}$ to 4 for $a/M=0.7$. This means that the detection rate of the coalescing Pop III BH-BHs ranges $8.3-7.2\times10^2\ {\rm events~yr^{-1} ~(SFR_p/(10^{-2.5}\msun~yr^{-1}~Mpc^{-3}))}$
${\rm \cdot([f_b/(1+f_b)]/0.33)}$.
The minimum detection rate of the coalescing Pop III BH-BHs corresponds to the worst model which we think  unlikely so that unless $ {\rm ~(SFR_p/(10^{-2.5}~\msun~yr^{-1}~Mpc^{-3}))\cdot([f_b/(1+f_b)]/0.33) \ll 0.1}$,  we can  expect the Pop III BH binary merger at least one event per year by the second generation gravitational wave detector. }

{
Fig. 3 of \cite{Nakano2015}  shows that if the S/N of the Pop III BH-BH QNM is 50, we can check the general relativity with the significance of  much more than 5 sigma level.
The criterion of S/N whether the significance has more than 5 sigma is  35.
Therefore, we expect the enough accuracy to discuss about the GR test with at least
one event of S/N= 35.
In our standard model, the detection rate of Pop III BH-BHs whose S/N is more than 35 is $3.2~\rm events~yr^{-1}$$(\rm SFR_p/(10^{-2.5}~\msun \rm~yr^{-1}~Mpc^{-3}))\cdot([f_b/(1+f_b)]/0.33)\cdot Err_{sys}$. 
Thus, there might be a good chance to check whether the GR is correct or not in the strong gravity region.
The forth and the seventh columns of Tables \ref{detecthu10035}-\ref{detectb14035}  show that there is a good chance to observe QNM of the merged BH since it reflects the space-time near the BH and their complex frequency does not depend on how it is excited.  The detection  of the expected QNM of the BH can confirm the GR in the strong gravity region. If it is different from the expected value, the true theory of gravity is different from GR.}

\begin{table*}
\caption{under100 cases with optimistic core-merger criterion, 1000 years, S/N $\geq$ 8}
\label{detecthu100}

This table shows the detection rates of Pop III BH-BHs for under100 cases with the optimistic core-merger criterion.
The first column
shows the name of the model. The second column shows the detection rate only by the inspiral chirp signal.
The third, the fourth and the fifth columns show the
detection rate only by the quasi normal mode (QNM) with Kerr parameter a/M = 0.70,
the detection rate by the quadrature sum of the inspiral chirp signal and the QNM with a/M = 0.70 and the detection rate by the linear sum of the inspiral chirp signal and the QNM with a/M = 0.70, respectively.
The sixth, the seventh and the eighth columns show the detection rates only by the QNM with a/M = 0.98, the detection rate by the quadrature sum of the inspiral chirp signal and the QNM with a/M = 0.98 and the detection rate by the linear sum of the inspiral chirp signal and the QNM with a/M = 0.98, respectively.
When signal-to-noise ratio of event that is calculated by matched filtering equation, 
over threshold S/N = 8, the event is detected.
All the rates are based on 1000 years Monte Carlo simulations.

\begin{center}
\footnotesize
\begin{tabular}{c c c c c c c c} 
\hline
14models  & \shortstack{Inspiral \\ S/N$ \geq 8$ \\ $\rm [/1000~yrs]$ } & 
            \shortstack{QNM(0.70)\\ S/N$ \geq 8$ \\ $\rm [/1000~yrs]$ } & 
            \shortstack{Quadrature sum of \\ Inspiral and \\ QNM(0.70) \\ S/N$ \geq 8$ \\ $\rm [/1000~yrs]$ } &
            \shortstack{Linear sum of \\ Inspiral and \\ QNM(0.70) \\ S/N$ \geq 8$ \\ $\rm [/1000~yrs]$ } &
\shortstack{QNM(0.98)\\ S/N$ \geq 8$ \\ $\rm [/1000~yrs]$ } & 
            \shortstack{Quadrature sum of \\ Inspiral and \\ QNM(0.98) \\ S/N$ \geq 8$ \\ $\rm [/1000~yrs]$ } &
           \shortstack{Linear sum of \\ Inspiral and \\ QNM(0.98) \\ S/N$ \geq 8$ \\ $\rm [/1000~yrs]$ } \\
\hline
our standard          &  85747&  67337& 180086&  392542  & 10680&   110896&  213115 \\
IMF:logflat          &  74764& 43130& 139705& 305428&  6524&    91378& 167958\\
IMF:Salpeter          &  47055&  16153& 74809& 156804&  2201&   53960&  90954\\
IEF:const.         &  80947&  65225& 173059& 380515         & 10476&  105409&205489\\
IEF:$e^{-0.5}$      &  77922&  63050& 167289& 369763& 10342&  101941&  198808\\
kick 100 $\rm km~s^{-1}$        &  66901&  57590& 148370&  330875&  9243&   88399& 176161\\
kick 300 $\rm km~s^{-1}$       &  16962&   14166&   36655& 80711&  2303&    22219&  43214\\
$\alpha \lambda$=0.01  &  18900&   13327&   37755&  82585&  2160&  23802&  44548\\
$\alpha \lambda$=0.1  &  67189&   73207& 165215& 366096& 11375&  93172& 193767\\
$\alpha \lambda$=10   &  74276&  59901&  156419&  336621& 10585&   96410&182207\\
$\beta$=0              &  85578&    67608& 180362&  393241& 10990&   110901&213343\\
$\beta$=0.5           & 117738&  74535&  229389& 498512& 11015&   145968& 272362\\
$\beta$=1              &  58823&  45926&  123468&  268790&  7353&   75646&145760\\
Worst                 &   4922&  2322&    8449&   17433&   316&     5786&   9997\\
\hline
\end{tabular}
\end{center}
\end{table*}

\begin{table*}
\caption{over100 cases with the optimistic core-merger criterion, 1000 years, S/N $\geq$ 8}
\label{detectho100}

The same as Table 18 but for over100 cases with the optimistic core-merger criterion.
\begin{center}
\footnotesize
\begin{tabular}{c c c c c c c c} 
\hline
14models  & \shortstack{Inspiral \\ S/N$ \geq 8$ \\ $\rm [/1000~yrs]$ } & 
            \shortstack{QNM(0.70)\\ S/N$ \geq 8$ \\ $\rm [/1000~yrs]$ } & 
            \shortstack{Quadrature sum of \\ Inspiral and \\ QNM(0.70) \\ S/N$ \geq 8$ \\ $\rm [/1000~yrs]$ } &
            \shortstack{Linear sum of \\ Inspiral and \\ QNM(0.70) \\ S/N$ \geq 8$ \\ $\rm [/1000~yrs]$ } &
\shortstack{QNM(0.98)\\ S/N$ \geq 8$ \\ $\rm [/1000~yrs]$ } & 
            \shortstack{Quadrature sum of \\ Inspiral and \\ QNM(0.98) \\ S/N$ \geq 8$ \\ $\rm [/1000~yrs]$ } &
           \shortstack{Linear sum of \\ Inspiral and \\ QNM(0.98) \\ S/N$ \geq 8$ \\ $\rm [/1000~yrs]$ } \\
\hline
our standard          &  84174&  218283&  332447& 611583&  41503& 142496& 275799\\
IMF:logflat          &  74305& 130893&   229061& 435444& 24041& 109695& 204510\\
IMF:Salpeter          &  47216&  39218&  98543& 191702&  6890&   59307& 101093\\
IEF:const.         &  79623&  222826& 332472& 609477& 42784& 139233& 271529\\
IEF:$e^{-0.5}$      &  77427& 221260& 327610& 598925& 43430& 136951& 266084\\
kick 100 $\rm km~s^{-1}$        &  66563&151555&  244415&473699&27246& 107999& 215263\\
kick 300 $\rm km~s^{-1}$       &  16843& 40830&  63663& 121315&7334&  27728&  54613 \\
$\alpha \lambda$=0.01  &  16864& 31946&  55038& 111547&5170&  25307&  49459\\
$\alpha \lambda$=0.1  &  62295&169260&257276&499826& 27164& 104855& 217745\\
$\alpha \lambda$=10   &  76876&181513& 284367&531845&34200&  125362& 240192\\
$\beta$=0              &  84596& 218993&333565& 613502&41842&  143224& 276753\\
$\beta$=0.5           & 118459& 98205& 255012&  540234&15457&  152208&  285242\\
$\beta$=1              &  58887& 46219& 123729& 268745&7482&  76071&145729\\
Worst                 &   4873& 2361&  8388&  17640&330&    5698&  9974 \\
\hline
\end{tabular}
\end{center}
\end{table*}

\begin{table*}
\caption{140 cases with the optimistic core-merger criterion, 1000 years, S/N $\geq$ 8}
\label{detecth140}

The same as Table 18 but for 140 cases with the optimistic core-merger criterion

\begin{center}
\footnotesize
\begin{tabular}{c c c c c c c c} 
\hline
14models  & \shortstack{Inspiral \\ S/N$ \geq 8$ \\ $\rm [/1000~yrs]$ } & 
            \shortstack{QNM(0.70)\\ S/N$ \geq 8$ \\ $\rm [/1000~yrs]$ } & 
            \shortstack{Quadrature sum of \\ Inspiral and \\ QNM(0.70) \\ S/N$ \geq 8$ \\ $\rm [/1000~yrs]$ } &
            \shortstack{Linear sum of \\ Inspiral and \\ QNM(0.70) \\ S/N$ \geq 8$ \\ $\rm [/1000~yrs]$ } &
\shortstack{QNM(0.98)\\ S/N$ \geq 8$ \\ $\rm [/1000~yrs]$ } & 
            \shortstack{Quadrature sum of \\ Inspiral and \\ QNM(0.98) \\ S/N$ \geq 8$ \\ $\rm [/1000~yrs]$ } &
           \shortstack{Linear sum of \\ Inspiral and \\ QNM(0.98) \\ S/N$ \geq 8$ \\ $\rm [/1000~yrs]$ } \\
\hline
our standard          & 57415&  526720& 605703&   880465 & 229397&301747& 468129\\
IMF:logflat          & 62822& 334314& 418581& 643524&135840& 210816& 336500 \\
IMF:Salpeter         & 45680&94589&153225&262258&34210& 85487&139481\\
IEF:const.         & 54288&517651& 593218&857973&223233& 292210& 453214\\
IEF:$e^{-0.5}$      & 52191& 510059&582984&840165& 217287& 283573&439967\\
kick 100 $\rm km~s^{-1}$       & 44081&501665&565394&816311& 181426&239150& 389811\\
kick 300 $\rm km~s^{-1}$       & 10582&126851& 142451&207343&56333& 70251& 108314\\
$\alpha \lambda$=0.01  &  9265&34743& 47728& 85457& 6163& 17465& 34075\\
$\alpha \lambda$=0.1  & 46225&318804& 387640&647921& 84572&146699& 280102\\
$\alpha \lambda$=10  & 46837& 394562&457263& 656125& 207658& 265181&389217\\
$\beta$=0              & 57037& 526812&606147& 881259&228443& 300762&467266\\
$\beta$=0.5           & 75890& 603167& 706657&1032803& 262573&  355852& 552073\\
$\beta$=1              & 34158& 256049& 304003& 460519&76025& 119493& 209309\\
Worst                 &  4101&  4262& 9803&  20095& 701&   5460&  10240\\
\hline
\end{tabular}
\end{center}
\end{table*}

\begin{table*}
\caption{under100 cases with the conservative core-merger criterion, 1000 years, S/N $\geq$ 8}
\label{detectbu100}

The same as Table 18 but for under100 cases with the conservative core-merger criterion

\begin{center}
\footnotesize
\begin{tabular}{c c c c c c c c} 
\hline
14models  & \shortstack{Inspiral \\ S/N$ \geq 8$ \\ $\rm [/1000~yrs]$ } & 
            \shortstack{QNM(0.70)\\ S/N$ \geq 8$ \\ $\rm [/1000~yrs]$ } & 
            \shortstack{Quadrature sum of \\ Inspiral and \\ QNM(0.70) \\ S/N$ \geq 8$ \\ $\rm [/1000~yrs]$ } &
            \shortstack{Linear sum of \\ Inspiral and \\ QNM(0.70) \\ S/N$ \geq 8$ \\ $\rm [/1000~yrs]$ } &
\shortstack{QNM(0.98)\\ S/N$ \geq 8$ \\ $\rm [/1000~yrs]$ } & 
            \shortstack{Quadrature sum of \\ Inspiral and \\ QNM(0.98) \\ S/N$ \geq 8$ \\ $\rm [/1000~yrs]$ } &
           \shortstack{Linear sum of \\ Inspiral and \\ QNM(0.98) \\ S/N$ \geq 8$ \\ $\rm [/1000~yrs]$ } \\
\hline
our standard          &  85078&67694& 179779& 392884&10940& 110427& 212302\\
IMF:logflat          &  75150& 43467& 140349&306831&6449&  91710&168361\\
IMF:Salpeter         &  47239&16174& 74706& 156639&2235& 54032&  90505\\
IEF:const.         &  81437& 65772& 173889&  382036&10614& 106027& 206030\\
IEF:$e^{-0.5}$      &  77642& 63262& 167088& 369913&10183& 101613& 198231\\
kick 100 $\rm km~s^{-1}$       &  67689& 58360&150101& 334191&9400&  89472&  177738\\
kick 300 $\rm km~s^{-1}$       &  17426& 14650&  37703& 83042&2355&22851&44532\\
$\alpha \lambda$=0.01  &  16187& 21897& 43744&  98315&4138&  23912& 50245\\
$\alpha \lambda$=0.1  &  72430& 64944& 162394& 361198&9863& 95990&192033\\
$\alpha \lambda$=10  &  74185&60266& 157326&337812&10294& 96638&182698\\
$\beta$=0              &  85590& 67951&  181242& 395187&10850& 110854& 214213 \\
$\beta$=0.5           & 117864& 74766& 229642& 498925&11113& 146208&272465\\
$\beta$=1              &  58961& 45682&123120&267733& 7462& 75987&145467\\
Worst                 &   4757& 2270& 8347& 17605&298&   5603& 10014\\
\hline
\end{tabular}
\end{center}
\end{table*}

\begin{table*}
\caption{over100 cases with the conservative core-merger criterion, 1000 years, S/N $\geq$ 8}
\label{detectbo100}

The same as Table 18 but for over100 cases with the conservative core-merger criterion

\begin{center}
\footnotesize
\begin{tabular}{c c c c c c c c} 
\hline
14models  & \shortstack{Inspiral \\ S/N$ \geq 8$ \\ $\rm [/1000~yrs]$ } & 
            \shortstack{QNM(0.70)\\ S/N$ \geq 8$ \\ $\rm [/1000~yrs]$ } & 
            \shortstack{Quadrature sum of \\ Inspiral and \\ QNM(0.70) \\ S/N$ \geq 8$ \\ $\rm [/1000~yrs]$ } &
            \shortstack{Linear sum of \\ Inspiral and \\ QNM(0.70) \\ S/N$ \geq 8$ \\ $\rm [/1000~yrs]$ } &
\shortstack{QNM(0.98)\\ S/N$ \geq 8$ \\ $\rm [/1000~yrs]$ } & 
            \shortstack{Quadrature sum of \\ Inspiral and \\ QNM(0.98) \\ S/N$ \geq 8$ \\ $\rm [/1000~yrs]$ } &
           \shortstack{Linear sum of \\ Inspiral and \\ QNM(0.98) \\ S/N$ \geq 8$ \\ $\rm [/1000~yrs]$ } \\
\hline
our standard          &  84864&214329& 329356& 608170&  41205& 142681&275555\\
IMF:logflat           &  75062& 128962& 227785&434489&23497& 110212& 205048\\
IMF:Salpeter         &  47104& 38994&  98198& 191541& 6847&  59142& 101187\\
IEF:const.          &  80521& 219273&330324& 606041&42209&139338& 271312\\
IEF:$e^{-0.5}$      &  77318&  218702&  325574& 595975& 42196& 135838&  265493\\
kick 100 $\rm km~s^{-1}$       &  67736&  150146& 243859&  474159&26286& 108305& 216333\\
kick 300 $\rm km~s^{-1}$       &  17139&   40582&  64028& 122219&  7137& 27864& 55062\\
$\alpha \lambda$=0.01  &  14194& 57924&  78827&  153302&9617&  28113&  59507\\
$\alpha \lambda$=0.1  &  66817&166466&259438& 497156&27926&109303& 219794\\
$\alpha \lambda$=10  &  76476&  182640& 285088& 532376&34371&125343& 240240\\
$\beta$=0              &  84415&  213727&328158& 607593& 40698& 141795& 274286\\
$\beta$=0.5           & 119110&  97920& 255288& 540990&15356& 152569&  286062\\
$\beta$=1              &  59051& 45600&  123150& 267169&  7314&  75779& 145191\\
Worst                 &   4819& 2337& 8335& 17592&333&  5702& 10007\\
\hline
\end{tabular}
\end{center}
\end{table*}

\begin{table*}
\caption{140 cases with the conservative core-merger criterion, 1000 years, S/N $\geq$ 8}
\label{detectb140}
The same as Table 18 but for 140 cases with the conservative core-merger criterion

\begin{center}
\footnotesize
\begin{tabular}{c c c c c c c c} 
\hline
14models  & \shortstack{Inspiral \\ S/N$ \geq 8$ \\ $\rm [/1000~yrs]$ } & 
            \shortstack{QNM(0.70)\\ S/N$ \geq 8$ \\ $\rm [/1000~yrs]$ } & 
            \shortstack{Quadrature sum of \\ Inspiral and \\ QNM(0.70) \\ S/N$ \geq 8$ \\ $\rm [/1000~yrs]$ } &
            \shortstack{Linear sum of \\ Inspiral and \\ QNM(0.70) \\ S/N$ \geq 8$ \\ $\rm [/1000~yrs]$ } &
\shortstack{QNM(0.98)\\ S/N$ \geq 8$ \\ $\rm [/1000~yrs]$ } & 
            \shortstack{Quadrature sum of \\ Inspiral and \\ QNM(0.98) \\ S/N$ \geq 8$ \\ $\rm [/1000~yrs]$ } &
           \shortstack{Linear sum of \\ Inspiral and \\ QNM(0.98) \\ S/N$ \geq 8$ \\ $\rm [/1000~yrs]$ } \\
\hline
our standard           & 55974&538382&616793& 893170&249666&321038& 489786\\
IMF:logflat           & 62224&  337179& 420100& 644264&146827&221150&346650\\
IMF:Salpeter        & 45352& 97882&155975& 265370&37519& 88623& 142980 \\
IEF:const.          & 53720& 528461&  603514&  869645&239380&307730& 470677\\
IEF:$e^{-0.5}$      & 51664&  515375& 587375& 845467&229457&  295251& 452615\\
kick 100 $\rm km~s^{-1}$      & 45815& 506994&  573221&830258& 184386& 244263&  399027\\
kick 300 $\rm km~s^{-1}$       & 11241& 135797& 152230&  219985&66444& 81421& 121657\\
$\alpha \lambda$=0.01  &  7431&   62666&   74189&  123411&14129& 24092&  47107\\
$\alpha \lambda$=0.1  & 47406&  366456& 437298&  710775&113247& 176176& 316298\\
$\alpha \lambda$=10  & 46402& 395992& 458184& 656940& 210422&267360& 391782\\
$\beta$=0              & 55773& 539238& 617239& 892843&249140& 320150& 489898\\
$\beta$=0.5            & 74932& 619682&722387& 1051940&277705& 370639& 570146\\
$\beta$=1              & 34014&   269616& 317328& 480619&79988&123652& 216096\\
Worst                &  4276&  4663&  10282& 20904&981& 5888& 10703\\
\hline
\end{tabular}
\end{center}
\end{table*}

\begin{table*}
\caption{under100 cases with optimistic core-merger criterion, 1000 years, S/N $\geq$ 35}
\label{detecthu10035}

This table shows the detection rates of Pop III BH-BHs for under100 cases with the optimistic core-merger criterion.
The first column
shows the name of the model. The second column shows the detection rate only by the inspiral chirp signal.
The third, the fourth and the fifth columns show the
detection rate only by the quasi normal mode (QNM) with Kerr parameter a/M = 0.70,
the detection rate by the quadrature sum of the inspiral chirp signal and the QNM with a/M = 0.70 and the detection rate by the linear sum of the inspiral chirp signal and the QNM with a/M = 0.70, respectively.
The sixth, the seventh and the eighth columns show the detection rates only by the QNM with a/M = 0.98, the detection rate by the quadrature sum of the inspiral chirp signal and the QNM with a/M = 0.98 and the detection rate by the linear sum of the inspiral chirp signal and the QNM with a/M = 0.98, respectively.
When signal-to-noise ratio of event that is calculated by matched filtering equation, 
over threshold S/N = 35, the event is detected.
All the rates are based on 1000 years Monte Carlo simulations.

\begin{center}
\footnotesize
\begin{tabular}{c c c c c c c c} 
\hline
14models  & \shortstack{Inspiral \\ S/N$ \geq 35$ \\ $\rm [/1000~yrs]$ } & 
            \shortstack{QNM(0.70)\\ S/N$ \geq 35$ \\ $\rm [/1000~yrs]$ } & 
            \shortstack{Quadrature sum of \\ Inspiral and \\ QNM(0.70) \\ S/N$ \geq 35$ \\ $\rm [/1000~yrs]$ } &
            \shortstack{Linear sum of \\ Inspiral and \\ QNM(0.70) \\ S/N$ \geq 35$ \\ $\rm [/1000~yrs]$ } &
\shortstack{QNM(0.98)\\ S/N$ \geq 35$ \\ $\rm [/1000~yrs]$ } & 
            \shortstack{Quadrature sum of \\ Inspiral and \\ QNM(0.98) \\ S/N$ \geq 35$ \\ $\rm [/1000~yrs]$ } &
           \shortstack{Linear sum of \\ Inspiral and \\ QNM(0.98) \\ S/N$ \geq 35$ \\ $\rm [/1000~yrs]$ } \\
\hline
our standard          & 2166&   488& 3172&  7190& 66& 2395& 4052 \\
IMF:logflat         & 1918&   326&  2591&  5544&48& 2062& 3292\\
IMF:Salpeter          & 1087&  102&  1353&    2719&13&1150& 1711\\
IEF:const.         & 2016&    476& 2945&   6679&62& 2238& 3732\\
IEF:$e^{-0.5}$      & 1937& 433&  2855&  6362&58&2156&3590\\
kick 100 $\rm km~s^{-1}$        & 1636&  404& 2456&   5623& 60&1850& 3123\\
kick 300 $\rm km~s^{-1}$       &  387& 91&  589&  1342& 9&  434& 732\\
$\alpha \lambda$=0.01  &  440&  96&   659&  1425&14&  483&  832\\
$\alpha \lambda$=0.1  & 1795& 520& 2882&6717&69&2050& 3569 \\
$\alpha \lambda$=10   & 1829& 436& 2733&5968&72& 2027&  3407\\
$\beta$=0              & 2118& 468&  3102& 7220&59& 2345& 4011\\
$\beta$=0.5           & 3176& 557& 4379& 9340&52& 3447& 5512\\
$\beta$=1              & 1574&  352& 2315& 5057&50& 1757&  2917\\
Worst          &  111& 21&  146&    297&      4&   118&   173 \\
\hline
\end{tabular}
\end{center}
\end{table*}

\begin{table*}
\caption{over100 cases with the optimistic core-merger criterion, 1000years, S/N $\geq$ 35}
\label{detectho10035}

The same as Table 24 but for over100 cases with the optimistic core-merger criterion.
\begin{center}
\footnotesize
\begin{tabular}{c c c c c c c c} 
\hline
14models  & \shortstack{Inspiral \\ S/N$ \geq 35$ \\ $\rm [/1000~yrs]$ } & 
            \shortstack{QNM(0.70)\\ S/N$ \geq 35$ \\ $\rm [/1000~yrs]$ } & 
            \shortstack{Quadrature sum of \\ Inspiral and \\ QNM(0.70) \\ S/N$ \geq 35$ \\ $\rm [/1000~yrs]$ } &
            \shortstack{Linear sum of \\ Inspiral and \\ QNM(0.70) \\ S/N$ \geq 35$ \\ $\rm [/1000~yrs]$ } &
\shortstack{QNM(0.98)\\ S/N$ \geq 35$ \\ $\rm [/1000~yrs]$ } & 
            \shortstack{Quadrature sum of \\ Inspiral and \\ QNM(0.98) \\ S/N$ \geq 35$ \\ $\rm [/1000~yrs]$ } &
           \shortstack{Linear sum of \\ Inspiral and \\ QNM(0.98) \\ S/N$ \geq 35$ \\ $\rm [/1000~yrs]$ } \\
\hline
our standard          & 2122& 959& 3707&  8269& 244&2587&4693\\
IMF:logflat          & 1866&  543& 2867&  6241& 117&2167& 3667\\
IMF:Salpeter          & 1098& 174&   1485&   2976&   36&1202& 1881\\
IEF:const.         & 1939& 953& 3475&  7756&276&2444& 4413\\
IEF:$e^{-0.5}$     & 1914&   928&3464& 7738&216& 2396& 4408\\
kick 100 $\rm km~s^{-1}$        & 1677&660& 2850& 6442&175& 2008& 3642\\
kick 300 $\rm km~s^{-1}$       &  382& 172&  668& 1512& 40&  472&  863 \\
$\alpha \lambda$=0.01  &  410&158&  689&  1454& 35&  479&  839\\
$\alpha \lambda$=0.1  & 1692& 780& 3054& 6900&159& 2066& 3923\\
$\alpha \lambda$=10   & 1762& 805& 3092&  6885&197& 2171& 3900\\
$\beta$=0              & 2247& 1002& 3808&  8397& 266& 2752& 4799 \\
$\beta$=0.5           & 2981&  637& 4349&  9394&95& 3328& 5569\\
$\beta$=1             & 1554&   341& 2294&  5071&53& 1753&  2909\\
Worst                 &  104&  13&  148&  331& 2&  116&   191 \\
\hline
\end{tabular}
\end{center}
\end{table*}

\begin{table*}
\caption{140 cases with the optimistic core-merger criterion, 1000 years, S/N $\geq$ 35}
\label{detecth14035}

The same as Table 24 but for 140 cases with the optimistic core-merger criterion

\begin{center}
\footnotesize
\begin{tabular}{c c c c c c c c} 
\hline
14models  & \shortstack{Inspiral \\ S/N$ \geq 35$ \\ $\rm [/1000~yrs]$ } & 
            \shortstack{QNM(0.70)\\ S/N$ \geq 35$ \\ $\rm [/1000~yrs]$ } & 
            \shortstack{Quadrature sum of \\ Inspiral and \\ QNM(0.70) \\ S/N$ \geq 35$ \\ $\rm [/1000~yrs]$ } &
            \shortstack{Linear sum of \\ Inspiral and \\ QNM(0.70) \\ S/N$ \geq 35$ \\ $\rm [/1000~yrs]$ } &
\shortstack{QNM(0.98)\\ S/N$ \geq 35$ \\ $\rm [/1000~yrs]$ } & 
            \shortstack{Quadrature sum of \\ Inspiral and \\ QNM(0.98) \\ S/N$ \geq 35$ \\ $\rm [/1000~yrs]$ } &
           \shortstack{Linear sum of \\ Inspiral and \\ QNM(0.98) \\ S/N$ \geq 35$ \\ $\rm [/1000~yrs]$ } \\
\hline
our standard        & 1410& 4487&  6553&  12149&1017& 2757&  5342\\
IMF:logflat          & 1580&  2585&4655&  8876&  572&2375& 4371\\
IMF:Salpeter         &  955& 624& 1823& 3652& 137& 1185&  2018 \\
IEF:const.         & 1387&4425& 6416& 11722&954& 2677& 5252\\
IEF:$e^{-0.5}$      & 1318&4220&6120& 11353&966& 2566& 5002\\
kick 100 $\rm km~s^{-1}$       & 1040& 2921& 4415& 8824&735&2028& 4072\\
kick 300 $\rm km~s^{-1}$       &  288& 968& 1370&  2489&206&   538& 1064\\
$\alpha \lambda$=0.01  &  236& 139&  436&   933& 44&  318&   556\\
$\alpha \lambda$=0.1  & 1187& 1606& 3331&  7358&397& 1823&  3789\\
$\alpha \lambda$=10  & 1210&  4366& 6072& 10415&949& 2418& 4404\\
$\beta$=0              & 1381& 4545& 6617&  12117&962& 2718&  5292\\
$\beta$=0.5           & 1904& 4858&  7457& 14128& 956& 3223& 6230\\
$\beta$=1              &  880& 1377&2633&  5571&393&1470& 2877\\
Worst                &   86&   19&  115&   248& 3&  99&  147\\
\hline
\end{tabular}
\end{center}
\end{table*}

\begin{table*}
\caption{under100 cases with the conservative core-merger criterion, 1000 years, S/N $\geq$ 35}
\label{detectbu10035}

The same as Table 24 but for under100 cases with the conservative core-merger criterion

\begin{center}
\footnotesize
\begin{tabular}{c c c c c c c c} 
\hline
14models  & \shortstack{Inspiral \\ S/N$ \geq 35$ \\ $\rm [/1000~yrs]$ } & 
            \shortstack{QNM(0.70)\\ S/N$ \geq 35$ \\ $\rm [/1000~yrs]$ } & 
            \shortstack{Quadrature sum of \\ Inspiral and \\ QNM(0.70) \\ S/N$ \geq 35$ \\ $\rm [/1000~yrs]$ } &
            \shortstack{Linear sum of \\ Inspiral and \\ QNM(0.70) \\ S/N$ \geq 35$ \\ $\rm [/1000~yrs]$ } &
\shortstack{QNM(0.98)\\ S/N$ \geq 35$ \\ $\rm [/1000~yrs]$ } & 
            \shortstack{Quadrature sum of \\ Inspiral and \\ QNM(0.98) \\ S/N$ \geq 35$ \\ $\rm [/1000~yrs]$ } &
           \shortstack{Linear sum of \\ Inspiral and \\ QNM(0.98) \\ S/N$ \geq 35$ \\ $\rm [/1000~yrs]$ } \\
\hline
our standard          & 2109&  498&3163&  7092& 54& 2350& 3986\\
IMF:logflat          & 1863&    302& 2646&   5530&40& 2034&  3307\\
IMF:Salpeter         & 1132& 132& 1421& 2755&17& 1200& 1804\\
IEF:const.         & 2012&  493& 3016&  6707&67& 2241&3816\\
IEF:$e^{-0.5}$      & 1905& 438& 2792&  6365&53& 2122& 3519\\
kick 100 $\rm km~s^{-1}$       & 1641&  384& 2492&  5613& 61&1825&3131\\
kick 300 $\rm km~s^{-1}$       &  376&  80& 565&   1278&11&  417&  720\\
$\alpha \lambda$=0.01  &  396& 164&  670& 1584&16&  467& 833\\
$\alpha \lambda$=0.1  & 1962&  525& 2939&  6624&69&  2191&  3714\\
$\alpha \lambda$=10 & 1757&   375& 2551&   5695&52& 1913&3193 \\
$\beta$=0              & 2221& 521& 3213&  7150& 64&2458& 4108\\
$\beta$=0.5           & 3168& 555& 4401& 9282&73&3454& 5532\\
$\beta$=1              & 1506& 330& 2223&  4925& 48&1675& 2843\\
Worst                 &   80&  13&  123&  301& 1&   89&  170\\
\hline
\end{tabular}
\end{center}
\end{table*}

\begin{table*}
\caption{over100 cases with the conservative core-merger criterion, 1000 years, S/N $\geq$ 35}
\label{detectbo10035}

The same as Table 24 but for over100 cases with the conservative core-merger criterion

\begin{center}
\footnotesize
\begin{tabular}{c c c c c c c c} 
\hline
14models  & \shortstack{Inspiral \\ S/N$ \geq 35$ \\ $\rm [/1000~yrs]$ } & 
            \shortstack{QNM(0.70)\\ S/N$ \geq 35$ \\ $\rm [/1000~yrs]$ } & 
            \shortstack{Quadrature sum of \\ Inspiral and \\ QNM(0.70) \\ S/N$ \geq 35$ \\ $\rm [/1000~yrs]$ } &
            \shortstack{Linear sum of \\ Inspiral and \\ QNM(0.70) \\ S/N$ \geq 35$ \\ $\rm [/1000~yrs]$ } &
\shortstack{QNM(0.98)\\ S/N$ \geq 35$ \\ $\rm [/1000~yrs]$ } & 
            \shortstack{Quadrature sum of \\ Inspiral and \\ QNM(0.98) \\ S/N$ \geq 35$ \\ $\rm [/1000~yrs]$ } &
           \shortstack{Linear sum of \\ Inspiral and \\ QNM(0.98) \\ S/N$ \geq 35$ \\ $\rm [/1000~yrs]$ } \\
\hline
our standard         &2103& 936& 3682&   8228&  243& 2543&4675\\
IMF:logflat           & 1918& 598&2945&  6221&155& 2231& 3697\\
IMF:Salpeter         & 1107& 210&  1483&  2977&   42&1215&1879\\
IEF:const.          & 1970& 884& 3410&  7700&  224& 2436& 4301\\
IEF:$e^{-0.5}$      & 1881&  922& 3364& 7568&  229&2315& 4240\\
kick 100 $\rm km~s^{-1}$   & 1592& 690&  2786&  6430&147& 1928&3578\\
kick 300 $\rm km~s^{-1}$    &  393&  172&  666&  1524&38& 479&  854\\
$\alpha \lambda$=0.01  &  385&  287&  783& 1715& 59& 509& 948\\
$\alpha \lambda$=0.1  & 1758& 752& 3030&   6862&183&2136& 3902\\
$\alpha \lambda$=10  & 1810& 834& 3184& 7080&216&2200& 4016\\
$\beta$=0              & 2157&  984& 3707&   8247& 248& 2653&  4649\\
$\beta$=0.5           & 3067&   677& 4413&  9573&71& 3402& 5587\\
$\beta$=1              & 1554&  336&2249&  4984& 43& 1715&  2843\\
Worst                 &  111&  18&  160&  358&4&  123&  205\\
\hline
\end{tabular}
\end{center}
\end{table*}

\begin{table*}
\caption{140 cases with the conservative core-merger criterion, 1000 years, S/N $\geq$ 35}
\label{detectb14035}
The same as Table 24 but for 140 cases with the conservative core-merger criterion

\begin{center}
\footnotesize
\begin{tabular}{c c c c c c c c} 
\hline
14models  & \shortstack{Inspiral \\ S/N$ \geq 35$ \\ $\rm [/1000~yrs]$ } & 
            \shortstack{QNM(0.70)\\ S/N$ \geq 35$ \\ $\rm [/1000~yrs]$ } & 
            \shortstack{Quadrature sum of \\ Inspiral and \\ QNM(0.70) \\ S/N$ \geq 35$ \\ $\rm [/1000~yrs]$ } &
            \shortstack{Linear sum of \\ Inspiral and \\ QNM(0.70) \\ S/N$ \geq 35$ \\ $\rm [/1000~yrs]$ } &
\shortstack{QNM(0.98)\\ S/N$ \geq 35$ \\ $\rm [/1000~yrs]$ } & 
            \shortstack{Quadrature sum of \\ Inspiral and \\ QNM(0.98) \\ S/N$ \geq 35$ \\ $\rm [/1000~yrs]$ } &
           \shortstack{Linear sum of \\ Inspiral and \\ QNM(0.98) \\ S/N$ \geq 35$ \\ $\rm [/1000~yrs]$ } \\
\hline
our standard        & 1373& 4983& 6979& 12584& 1140&2858& 5385\\
IMF:logflat           & 1548& 2862& 4879&   9012&614& 2412&  4321\\
IMF:Salpeter       & 1030& 714&1977& 3769& 153&   1263& 2088\\
IEF:const.          & 1336& 4820& 6796&12216&1095& 2714& 5218\\
IEF:$e^{-0.5}$      & 1354& 4614& 6574& 11850&1029&2694& 5120\\
kick 100 $\rm km~s^{-1}$   & 1144& 2951&  4591& 9106&760&  2136&  4157\\
kick 300 $\rm km~s^{-1}$   &  313& 1192&  1616&   2749&284&  643&  1185\\
$\alpha \lambda$=0.01  &  170&  244& 480& 1053&  71&  289&  588\\
$\alpha \lambda$=0.1  & 1167&   2007& 3695&  7679&522&  1924& 3901\\
$\alpha \lambda$=10  & 1144&  4244& 5915& 10287&854& 2270& 4246\\
$\beta$=0              & 1432& 5100&7136&  12715& 1189& 2921&  5482\\
$\beta$=0.5            & 1786&   5209& 7792&  14520&1085&3206& 6250\\
$\beta$=1              &  811&   1424& 2596&  5658&430&1407& 2853\\
Worst                &   99&   35&  151&   313&2&  111&   182\\
\hline
\end{tabular}
\end{center}
\end{table*}

\section{Discussion \& Summary}\label{discussion}
In this paper, we performed the Pop III binary population synthesis and examined the parameter dependence of Pop III binary evolutions.
We examined  the dependence of the results on IMF, IEF, the natal kick velocity, the CE parameters and the lose fraction of  stellar mass.
As for the chirp mass distribution, each model has the peak at around $30 ~\msun$.
In  several models, the chirp mass distribution has a tail from $30~\msun$ to more massive region. However the robust property is that the chirp mass distribution has the peak at $30~\msun$. 

In order to compare the variability of $\rm Err_{sys}$, we refer previous researches such as \cite{Belczynski2012a,Dominik2015}.
In \cite{Belczynski2012a}, they calculated the solar metal ($Z=0.02$) binaries and 10\% solar metal ($Z=0.002$) binaries to estimate the detection rates assuming  as  half of the stars formed with solar metal and the other with10\% solar metal.
In \cite{Belczynski2012a}, they calculated 20 models by varying the maximum NS mass, the natal kick velocity, rapid or delayed supernova models  which  change the mass spectrum of supernova remnants, wind mass loss and  $\beta$. They also considered whether in the Hertzsprung gap donors always merge companion during the CE phase or not. 
The detection rate of their realistic Standard model in \cite{Belczynski2012a} is 517.3 $\rm events~yr^{-1}$
The detection rates of \cite{Belczynski2012a} are from $14~\rm events~yr^{-1}$ to $12434.4~\rm events~yr^{-1}$.
Thus, $\rm Err_{sys}$ of \cite{Belczynski2012a} is from $2.7\times10^{-2}$ to 24.

On the other hand, \cite{Dominik2015} calculated binaries whose metallicity range is from $Z=10^{-4}$ to $Z=0.03$ and estimated the detection rate using the metallicity and SFR evolution models. 
There are 16 models  varying high-end or low-end metallicity models, whether in the Hertzsprung gap donors always merge companion during the  CE phase or not, rapid or delayed supernova models, the natal kick and waveform models.
The detection rate of their Standard model of high-end metallicity scenario in \cite{Dominik2015} is 306 $\rm events~yr^{-1}$
The detection rate of \cite{Dominik2015} is from $8.2~\rm events~ yr^{-1}$ to $3087~\rm events~yr^{-1}$ by the 3-detector network using inspiral and PhC waveform (S/N=10).
Thus, $\rm Err_{sys}$ of \cite{Dominik2015} is from $2.7\times10^{-2}$ to 10.
Therefore, the variability of $\rm Err_{sys}$ of Pop III is less than that of Pop I and Pop II, although the models are different.
There are two reasons for this difference.
Firstly,  Pop III star binaries do not enter the CE phase so that the result does not depend on the treatment of the CE phase so much.
Secondly, the Pop III compact binaries are more massive than the Pop I compact binary so that the Pop III binaries are hard to be disrupted by the natal kick.
Therefore, the property of the chirp mass distribution and the detection rate are likely robust result.
{However, note that in the case of the detection rate of Pop III there are the dependence on the SFR and $\rm f_b$ yet.}

There are some uncertainties yet such as the separation distribution function and the mass ratio distribution function for which we did not alter. The former  will change the number of close binaries which can have  binary interactions.
Therefore this effect may change the event rate, but the property of chirp mass distribution is not likely changed a lot because the binary interaction is not changed. 
From our Monte Calro simulations, the chirp mass distribution of Pop III BH-BHs is upward to the high mass and has a peak at  
$\sim 30\ \msun$ in each model.
The compact objects in IC10 X-1 and NGC300 X-1 may be around 30 $\rm M_{\odot}$ and they might become coalescing massive BH-BHs whose chirp masses are 11-26 $\rm M_{\odot}$ (See \cite{Bulik2011}).
Thus, Pop I stars or Pop II stars might become coalescing massive BH-BHs.
However, the observed typical mass of Pop I BH-BHs is around $10~\rm M_{\odot}$ and massive BH like IC10 X-1 and NGC300 X-1 would be rare  (See also Fig. 1 in \cite{Belczynski2012b}) so that the chirp mass distribution of Pop I BH-BHs might be flat or decreasing as a function of mass.
{The result of the binary population synthesis simulation for Pop I and Pop II stars by \cite{Dominik2015} also suggests that the chirp mass distribution of Pop I BH-BHs might be flat or downward to high mass (See the Fig. 7 in \cite{Dominik2015}).
Furthermore, the Pop I and Pop II BH-BH detection rate of the standard model in \cite{Dominik2015} is 306 $\rm yr^{-1}$.
The fraction of the Pop I and Pop II BH-BH whose mass is larger than $20~\msun$ is about 25\%.
Thus, the detection rate of the Pop I and Pop II high mass BH-BHs is expected as about $80~\rm yr^{-1}$.
Note that this value depends on the $\rm Err_{sys}$ of \cite{Dominik2015} which is from $2.7\times10^{-2}$ to 10.
Therefore, if the detection rate of the coalescing Pop I and Pop II high mass BH-BHs is lower than that of Pop III, we may be able to confirm the existence of Pop III star by the detection of the chirp signal 
and QNM to determine the chirp mass and the total mass distribution since the typical mass of Pop III BH-BH binary is much larger than those of observed Pop I BH.
On the other hand, if the detection rate of the coalescing
Pop I and Pop II high mass BH-BHs is higher, Pop III BH-BHs contribute only some parts of the gravitational wave events of BH-BHs.
In this case, the existence of Pop III binaries will be confirmed by the investigation of the merger rate history as function of redshift 
by  DECIGO (DECi hertz Interferometer Gravitational wave Observatory) \citep{seto2001}. }

As for the mass ratio distribution function, if the number of the high mass ratio (i.e. near 1) increases, the number of BH-BH probably increase. On the other hand, if the number of the low mass ratio increase, the number of BH-BH will decrease while the number of NS-BH will increase. We will check the dependence of these two initial distribution functions in future work.  
Development in the simulation may make it possible to clarify  initial conditions of Pop III binary.
 
The Pop III star formation rate will determine  the merger rate.
{Our using Pop III SFR \citep{de Souza2011} is Fig. \ref{sfr}.
}
There are some arguments on  Pop III SFR besides \cite{de Souza2011}. 
For example, \cite{Johnson2013} simulated the Pop III SFR by the smooth particle hydrodynamics (SPH) simulations.
In their simulations, the peak value of Pop III SFR is from $\sim10^{-3.7}~\msun \rm~yr^{-1}~Mpc^{-3}$ to $\sim10^{-3}~\msun \rm~yr^{-1}~Mpc^{-3}$ at $z\sim10$. 
The difference between the high value and the low value comes from without Lyman-Werner (LW) feedback or with LW feedback. Note that the result of these simulations might change if the metal pollution model changes.  
On the other hand, \cite{Kulkarni2014} and \cite{Yajima2015} studied the Pop III SFR by considering the contribution of Pop III stars to cosmic reionization. 
\cite{Kulkarni2014} suggest that the peak value of Pop III SFR is from $\sim10^{-4.2}~\msun \rm~yr^{-1}~Mpc^{-3}$ to $\sim10^{-1.3}~\msun \rm~yr^{-1}~Mpc^{-3}$ at $z\sim10$.
The difference between the high value and the low value comes from  that of the metal pollution timescale.
While \cite{Yajima2015} suggests that the peak value of Pop III SFR is $\sim10^{-3}~\msun \rm~yr^{-1}~Mpc^{-3}$ at $z\sim15$ in order to recover the observed Thomson scattering optical depth of the cosmic microwave background.
The SFR of Pop III is  controversial now.
However, these estimated value of the SFR tell us that except for the worst model, we might expect the detection of GW from massive Pop III BH-BH near future. 

{The present merger rate density which is calculated only from SFR between z=7 to z= 11 is about 50\% of the whole merger rate density.
Therefore $\rm SFR_{peak}$  is a good parameter in our adopted model of SFR.
However, there is Pop III SFR whose peak region is higher redshift than $z\sim10$ such as the models in Yajima et al. (2015). 
In such a  case, we have to consider $\int SFR(z)dz $ in order to compare the dependence on the SFR.}

We discuss also the binary fraction of Pop III.
Recently, the resolution of the multi-dimension simulation becomes so high that the fragmentation of disk at the Pop III stellar formation can be studied like in \citep{Clark2011}.
The recent cosmological hydrodynamics simulation \citep{Susa2014} suggests that the binary fraction is about 50\%.
Thus, we use $\rm f_b=1/2$ since the total number of the binary is half of the total number of the stars in the binary.
However, the binary fraction is controversial.
Thus, we express the uncertainty of the binary fraction of Pop III as $\rm f_b$. 

From Fig. \ref{mergerz}, the peak of the rate of the  merger of Pop III BH binary is around $z=9$ so that
the observed frequency of the chirp signal and the quasi normal mode are $\sim$ 10 times small. To detect
such a low frequency gravitational wave, DECIGO
\citep{seto2001} will be most appropriate.
When DECIGO starts an observation around 2030, we can detect gravitational waves from Pop III BH-BHs which merged at $z\sim 9$. Thus, we might identify  the peak of Fig. \ref{mergerz}.
The peak of Fig. \ref{mergerz} depends not only on binary parameters, but also on the Pop III SFR.
Therefore, we might get the information of Pop III SFR.

In this paper, the merging NS-NS and NS-BH are not considered because they are negligibly small in number  in almost all models. However, in some models they are not so. Since they are the candidates of the short gamma ray burst (GRB),
 the high redshift observation of GRB by  Hi-z GUNDAM \citep{Yonetoku2014} might be possible.
Thus, the merging NS-NS and NS-BH might also be useful for Pop III binary parameter studying.

\section*{Acknowledgements}
 We thank Nakauchi for useful comments at the early phase of the draft.
We also thank Nakano for useful comment and discussion.
 This work is supported in part by the Grant-in-Aid from the
Ministry of Education, Culture, Sports, Science and Technology (MEXT)
of Japan, Nos. 251284 (TK),  23540305, 24103006,15H02087(TN), 24103005(NK).

\bibliographystyle{mn2e}

\end{document}